\documentclass[%
showkeys
]{revtex4-2}
\usepackage{graphicx}
\usepackage{dcolumn}
\usepackage{bm}

\usepackage{float}
\usepackage{subcaption}
\usepackage{siunitx} 
\usepackage[cm]{fullpage}
\usepackage{multirow}
\usepackage[utf8]{inputenc}
\usepackage{array}
\usepackage{enumerate}
\usepackage{hyperref}
\usepackage{xfp}
\usepackage{tabularx}
\usepackage{booktabs}

\newcolumntype{L}[1]{>{\raggedright\let\newline\\\arraybackslash\hspace{0pt}}m{#1}}

\newcommand{\beq}{\begin{equation}}
\newcommand{\eeq}{\end{equation}}
\newcommand{\bea}{\begin{eqnarray}}
\newcommand{\eea}{\end{eqnarray}}

\begin{document}

\title[2D Trypanosoma cruzi laminar flow model]{Mechanical Evidence of the impossibility of directed motion of \emph{Trypanosoma cruzi} towards preferred organs in the Human Body, a simulation 2D model within a laminar flow}

\author{Alberto-Mario Castillo}

\affiliation{
Bogotá, Colombia\\
\url{https://orcid.org/0009-0004-3024-3449}
}

\author{Gabriel Villalobos}
\affiliation{
Escuela Superior de Administración Pública\\
Facultad de Posgrados, Bogotá, Colombia \\
\url{gabriel.villalobos@esap.edu.co}\\
\url{https://orcid.org/0000-0003-4582-265X}
}

\date{\today}


\begin{abstract}
The movement of the infective form of the T. Cruzi parasite within the
human blood is not completely understood.  Video microscopy
observations confirm forward motility of the protozoa and relate it to
the deformation of the body, nonetheless there are open questions
relating the deformation of the protozoan with its motion in blood,
for which a computational model would be very helpful. Hereby 
we introduce a simple computational 2D model to test whether it is
mechanistically possible for the parasite to direct its movement
through the blood stream towards preferred organs in the bloodstream.

We model the infective form of the T. cruzi, the causative agent of
Chagas disease; by means of a network of harmonic springs that
represents its body, without explicitly modeling the flagellum. We
coupled this with a particle model of a laminar fluid flow, which we
implemented using the Dissipative Particles Dynamics method.

Our parasite model swims on still fluid, its center of mass
displacement being larger than the end to end deformation. Laminar
flow dragged the model parasite of the order of 5 $\mu m$; having
minor effects on its structure.

The parasite model does not exhibit directed motion in a moving fluid,
suggesting that the large-scale displacement of the real parasite is
not attributable to its own motility but is instead caused by it being
dragged by the flow. We highlight two potential improvements for the
model: incorporating the capacity to describe 3D motion and including
a specific depiction of the flagellum.
\end{abstract}


\maketitle                                                                                                                                                             
 \newpage


\section{Introduction}

Chagas disease or American Trypanosomiasis is caused by \emph{Trypanosoma cruzi (T. Cruzi)}, a protozoan parasite that flows within blood; while living and reproducing inside its human host for several years would cause different diseases and even death\cite{villalta2009perspectives}. The parasite can be transmitted to humans not only by blood-sucking insects but through organ transplantation or congenital transmission\cite{stanaway2015burden}. Due to migration phenomena and global trading, Chagas has spread around the world, leading to a global infected population of almost eight million \cite{world2012research}
The annual death toll of this neglected tropical disease is close to $12500$ people, making Chagas the parasitic disease in Latin America with the higher socioeconomic impact \cite{world2012research}. Such recognition not only obeys to cardiovascular morbidity but to premature death of patients that are economically productive, aged in the range of 20 to 50 years.

The dynamics of the protozoan within the blood flow, while relevant to the development of early stages of the disease, has not been completely understood. Recent research, including studies on parasite motility and invasion mechanisms \cite{arias2020motility}, has provided valuable insights, yet these efforts remain insufficient to fully elucidate the intricate behaviors and interactions within the bloodstream.
It has been reported that the protozoan shows forward parasite motility related to the beating of the flagellum \cite{Ballesteros12}. Nonetheless, it is crucial to have a more detailed description that can be used to test different hypotheses. Firstly, in blood there are corpuscles that are similar in size to the protozoan. Secondly, the observations are unable to test the relative importance of the deformation of the protozoan body itself to the beating of the flagellum.
This could be accomplished by having a computational model system, in which such measurements are easily implemented, and thus those hypotheses are tested.

In this work we present a simple 2D model of the protozoan within a basic fluid model. We then assess whether such a system shows tropism and directed motion. This model can be used as a tool to test hypothesis of the conditions under which the real parasite could show those traits.
There are many advantages in using a simple model to try to capture the motility of the protozoan, among them the possibility of running such system less expensively. Therefore, we find it useful to test the possibility of having a 2D model that does not include the flagellum as a first attempt to capture the dynamics. 
Nonetheless, we understand that this is a very simple model, and further work is needed to capture the 3D structure of the parasite and to understand the interaction with blood organelles as red blood cells in a more detailed model.

The rest of the paper is structured as follows. 
In section \ref{sec:model} we present the elastic model for the body of the protozoan.
We take into account the deformation of the protozoan by discretizing its shape from the literature \cite{finkelsztein2015altering} and having our model switch among them.
Furthermore, we use the dissipative particle model (DPD from now on), for the representation of the fluid. Finally, we couple both the fluid model and the body. This modular design would allow for increasing the complexity of the model in the future, for instance having other corpuscles, as red blood cells.
Afterward, in section \ref{sec:numexp} we show our results for the movement of the model parasite in different flow conditions. We successfully characterize the system that accounts for the interaction between the fluid flow and the protozoan model. Nonetheless, we do not find directed motion as seen in the descriptive studies from the literature. 
Finally, in section \ref{sec:conclusion} we present our final remarks and comments.

\section{Model} \label{sec:model}
\subsection{Fluid model}
The chosen hydrodynamic model is Dissipative Particle Dynamics, as introduced by \cite{Hoogerbrugge_1992}. The fluid is represented by \(N\) particles, whose kinematic variables are  positions \(r_i\) and momenta \(p_i\). Movement of the particles happens in discrete intervals of time \(\delta t\) in two phases. First, a collision phase which updates the momentum of the particles within the fluid:
\begin{equation}
  {p_i}' = p_i+\sum_j\Omega_{ij}\textbf{e}_{ij}\label{eq:Collision}
\end{equation}

\noindent \(\Omega_{ij}\) represents the interchange of momentum of the particles \textit{i} and \textit{j}. It results from adding up the conservative, dissipative, and random forces: 
\begin{equation}
  \Omega_{ij} = (F_{ij}^{C}dt + F_{ij}^{D}dt + F_{ij}^{R}\sqrt{dt})e_{ij}
  \label{Omegaij}
 \end{equation}

The interaction between particles depends on its distance, and leads to an equilibrium between the stochastic term and the relaxation of the relative motion as described in \cite{nikunen2003would}.

The second phase is called streaming. The position of the particles is updated by displacing them along the direction of the momentum during the time $\delta t$:
\begin{equation}
  r_i' = r_i+\frac{\delta t}{m_i}p_i'\label{eq:Propagation}
\end{equation}

Unless stated otherwise, we apply periodic boundary conditions in the horizontal direction and bounce back conditions on the vertical limits of the simulation box \cite{REVENGA1999309}.

\subsection{T. Cruzi parasite model}
We propose to model the body of the protozoan as a flexible structure, following similar modeling of biological structures in the literature \cite{mcwhirter2009flow} under the observation that both red blood cells and T. cruzi share similar microscopic scales and can both be conceptualized as fluid-filled vesicles exhibiting physical properties, such as elasticity. To do so, we discretize the surface membrane of the protozoan by a set of node particles \(N_{s}\) that are arranged in a two-dimensional triangular mesh with $N_t$ triangles.

Related to this mesh there are three elastic components. Firstly, each link of the mesh is regarded as a spring, therefore there is a elastic energy related to a rest length. Secondly, each triangle has an elastic energy associated with its area. Thirdly, there is a bending energy associated with the conservation of the angles within each triangle.
\begin{equation}
  \label{eq:energiaparasito}
  U = U_m + U_b + U_A
\end{equation}
\noindent Where \(U_m\) is the potential energy stored in the elongation of the springs, \(U_b\) the potential bending energy, and \(U_A\) is the potential energy stored in the area \cite{fedosov2011multiscale}.

\[\label{eq:potencialelastica}
U_b = \sum_{j \in N_s}k_b\left( 1- \cos\left(\theta_j-\theta_{j0}\right)\right)
\]

\[\label{eq:potencialarea}
U_A = \sum_{j \in N_t}k_d\frac{\left( A_j-A_{j0}\right)^2}{2A_0}
\]

\[\label{eq:potencialelastica}
U_m = \sum_{j \in N_s}k_m \frac{\left( l_j-l_{j0}\right)^2}{2}
\]

As reported in \cite{finkelsztein2015altering}, the parasite continuously deforms as it moves. To capture this dynamic behavior, we constructed five distinct discretizations based on video recordings from \cite{finkelsztein2015altering} (Table \ref{tab:Discret}).

\begin{table}[H]
  \centering
    \caption [Discretizations triangular meshes]{Five mesh discretizations used in this work.}
  \begin{tabular}{|l|l|l|l|l|}\hline
    \includegraphics[width = 0.15\columnwidth]{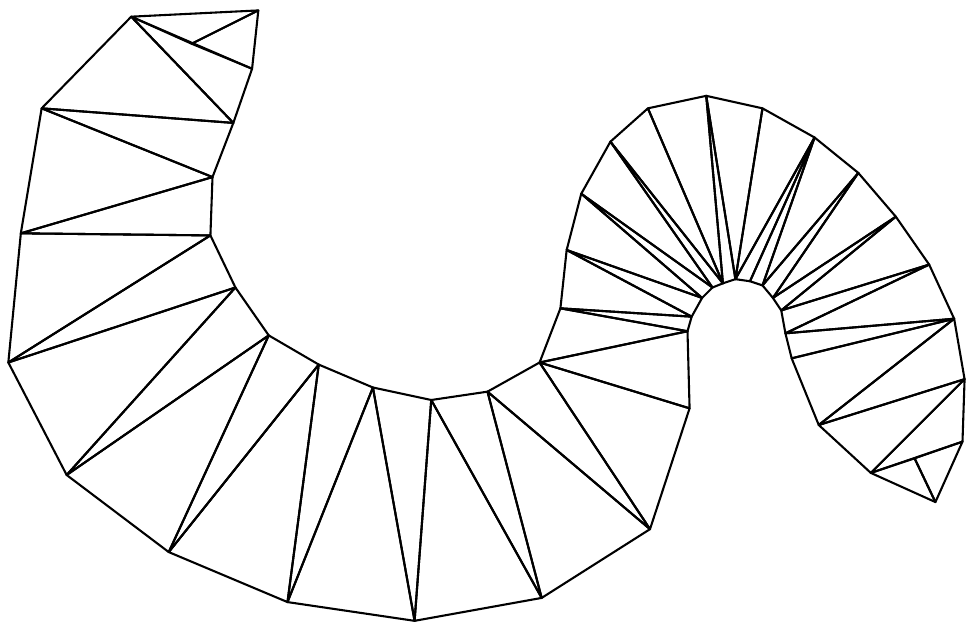} & \includegraphics[width = 0.15\columnwidth]{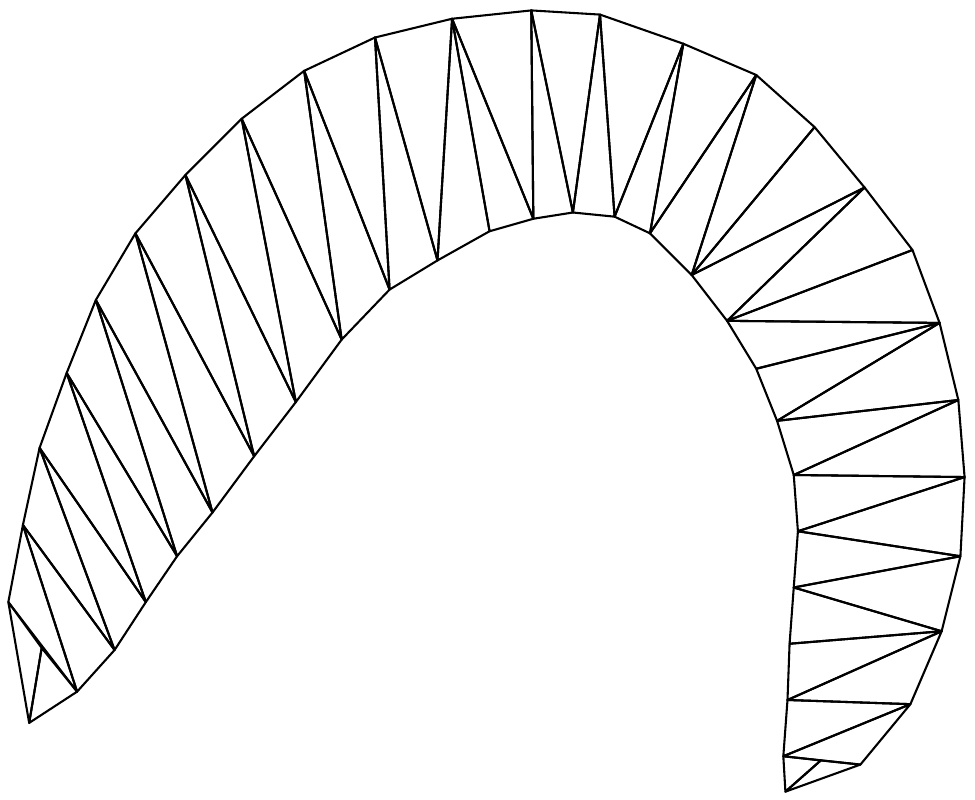} &  \includegraphics[width = 0.15\columnwidth]{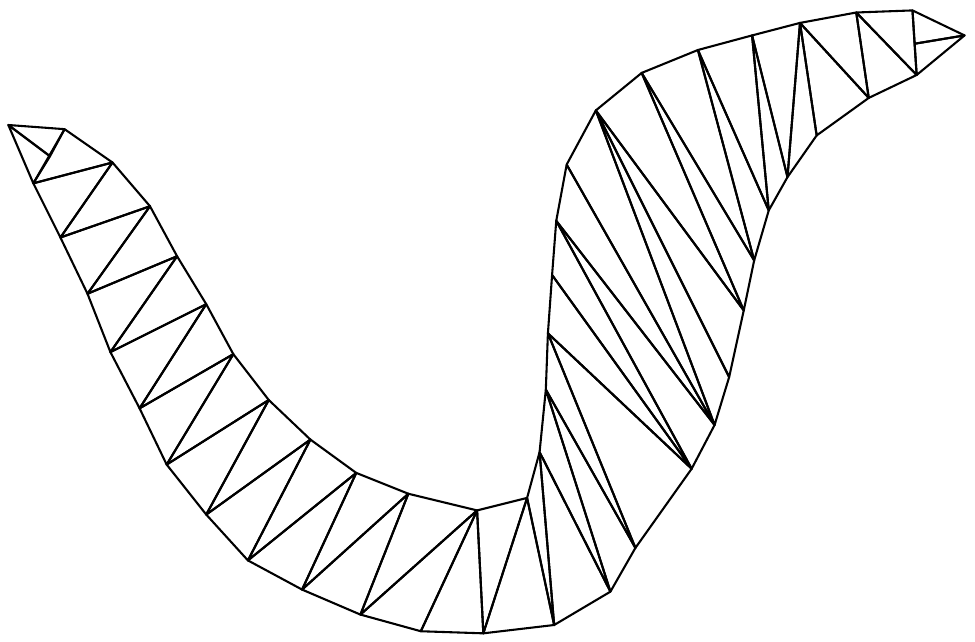} & \includegraphics[width = 0.15\columnwidth]{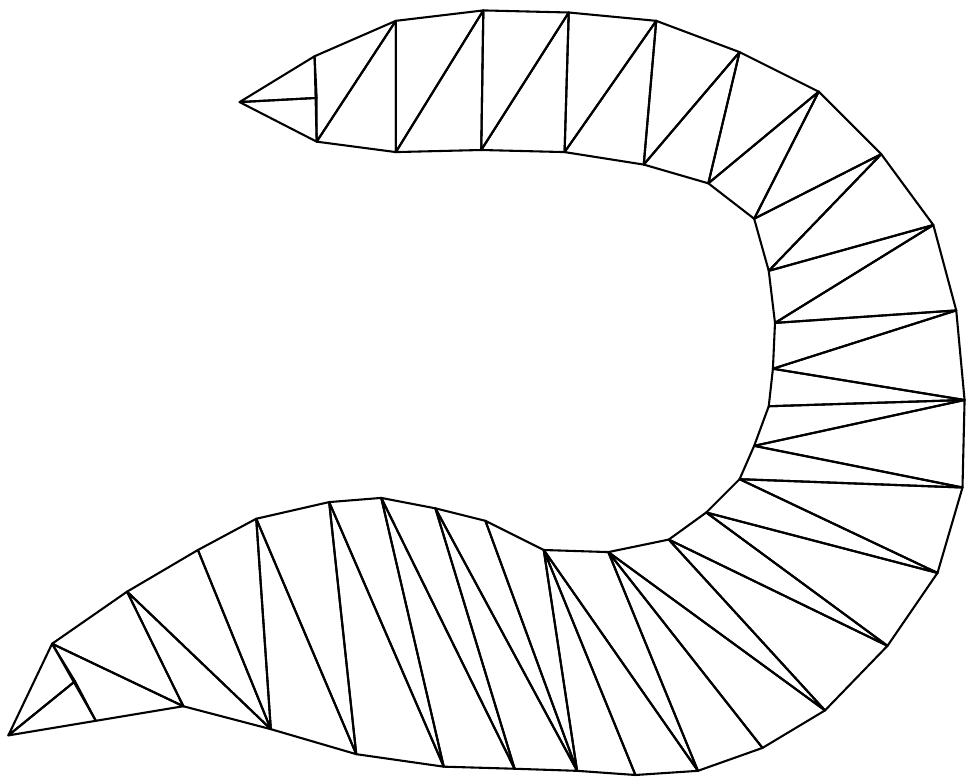} & \includegraphics[width = 0.15\columnwidth]{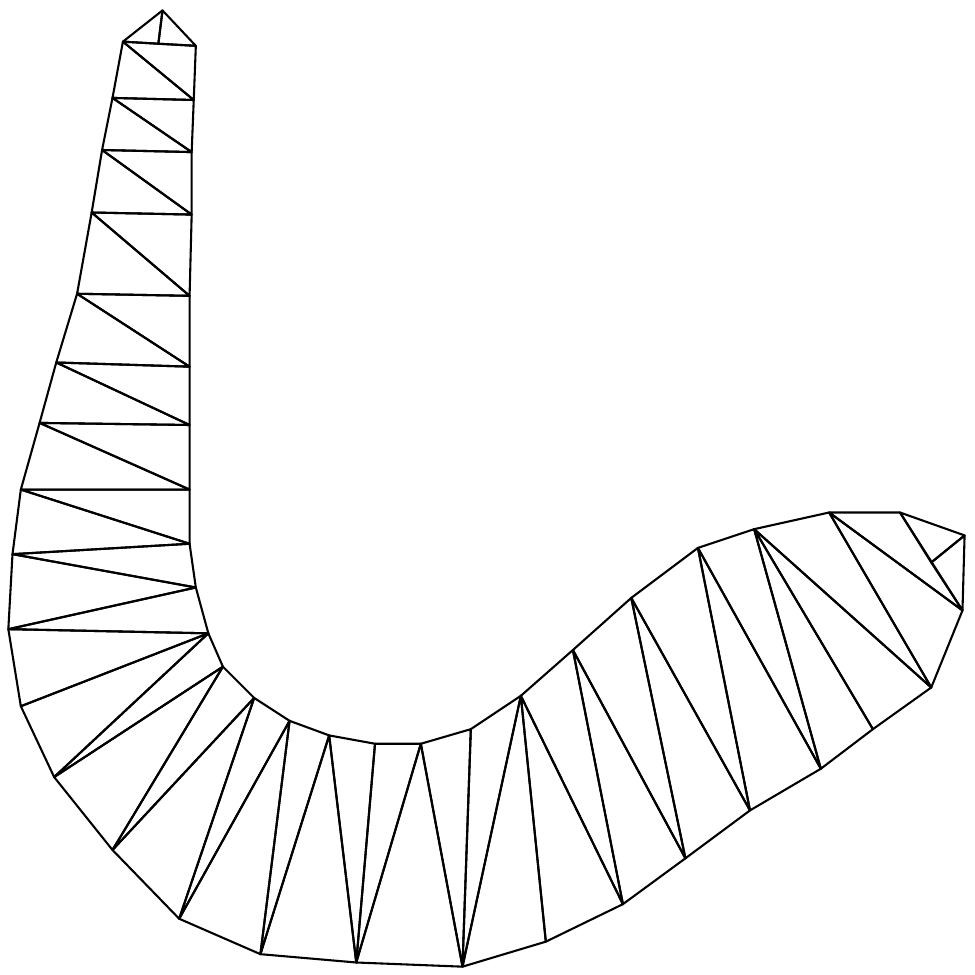} \\ \hline
  \end{tabular}
  \label{tab:Discret}
\end{table}

Each discretization $k$ has a unique set of rest parameters: for each angle $\theta_j$ a rest angle $\theta_{j0}$; for each triangle, a rest area \(A_{j0}\); and for each spring, a rest length \(l_{j0}\).
Since the parasite changes its shape as a function of time, the model simulates locomotion by sequentially transitioning between the discretizations. Each transition is achieved by updating the model's target rest parameters \(\{l_{j0}\},\{A_{j0}\}\) and \(\{\theta_{j0}\}\) to match those of the next discretization in the sequence. Then, over a series of discrete time steps, the model's instantaneous shape —defined by its current lengths \(\{l_{j}\}\), areas \(\{A_{j}\}\), and angles \(\{\theta_{j}\}\)— smoothly evolves until it settles into this new equilibrium defined by the target parameters.


The authors acknowledge the flagellum plays a crucial role in the dynamics and locomotion of the protozoan \cite{arias2020motility}, its effect is implicitly incorporated into the model through the discretization of the movement. This approach captures the overall behavior of the protozoan body without explicitly modeling the detailed interactions of the flagellum. This simplification allows us to establish a preliminary framework to evaluate the model and its potential extensions for capturing more complex dynamics in future studies.


\subsection{Parasite-Fluid Interaction}\label{sec:parasite-fluid-interaction}
We model the interaction between the parasite and the DPD particles as a bounce back collision on the links that correspond to the border of the parasite. The elastic collision would result in a force that is applied to the two nodes that are joined by the corresponding link, \(\textbf{F}_{i} = \frac{1}{2}\frac{\Delta \textbf{p}}{\Delta t}\); where \(\textbf{p}\) is the momentum transfer of the DPD particle and \(\Delta t\) the collision time.
For particles crossing the structure limits, their relative position to the outermost node is calculated, and a force is derived from it as:
\begin{eqnarray}
  \Delta \textbf{p} &=&  \iota \times \kappa \times \vec{\rho} \times \Delta t \label{Interaction Momentum}
\end{eqnarray}

Here $\Delta \textbf{p}$ is the momentum change, $\iota$ represents how deep a particle went into a triangle, $\kappa$ accounts for the strength of the interaction, and $\vec{\rho}$ is a vector normal to the $Z$ axis pointing out the cell body:

\begin{figure}[H]
  \centering
  \includegraphics[width = 0.3\textwidth]{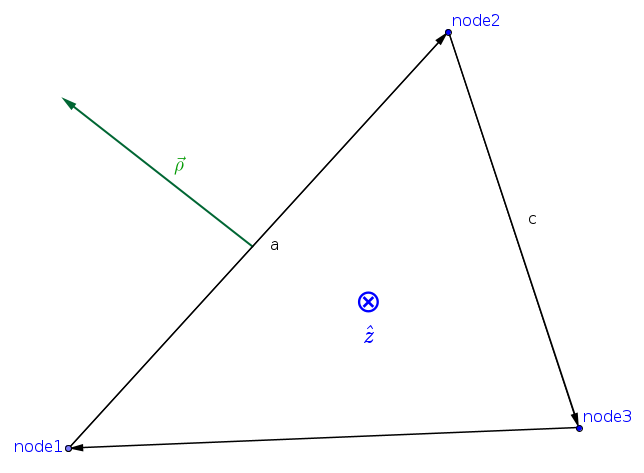}
  \caption{Collision stage. Momentum vector pointing out of the structure.}
  \label{fig:collision-points-out}
\end{figure}

The vector $\vec{\rho}$ plays a key role in the coupled performance of the models, as its primary function is to keep fluid particles outside the structure. This is achieved by performing a check for each triangle of the mesh. For any given triangle and a nearby fluid particle, the dot product is calculated between the triangle's outward normal vector and the particle's position vector relative to the two nodes that form the crossed link. The sign of the dot product indicates whether these vectors point in the same or opposite directions. If the directions are opposite, the particle is marked as crossing the structure's limits.


\begin{figure}[H]
  \centering
  \includegraphics[width = 0.5\textwidth]{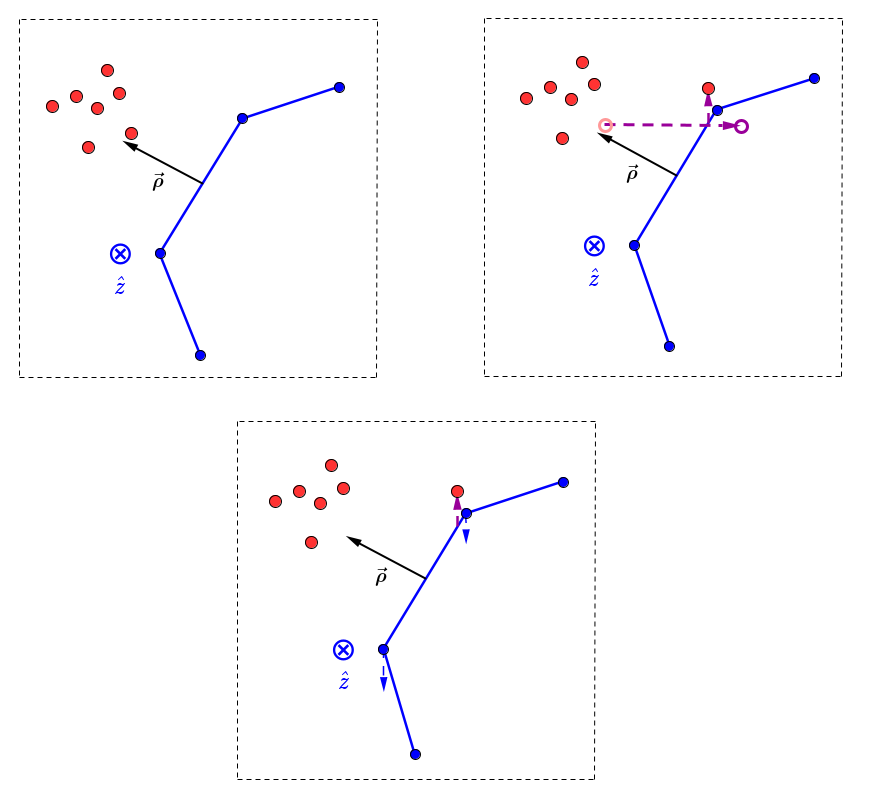}
  \caption{Collision stage. Group of fluid particles approaching to cell membrane.}
  \label{fig:collision-bounce-back}
\end{figure}

When a particle crosses the limits of the structure, its relative position with respect to the outermost node is calculated, and a force is derived from it. The resulting momentum updates the particle's position in the perpendicular direction and is also transferred to the structure's nodes as a force.

In this way, particles as well as nodes —and in consequence the triangles, will get an update in their positions.

\section{Model Units} \label{sec:munits}
The model internal units refer to a system of units used within the simulation and are needed to parameterize the real system into it, nonetheless not all the required information is available. Some of the values were estimated, others freely chosen, and a third set of parameters depend on the first two sets and therefore, were calculated.

The model length unit $(l)$ was set to be equivalent to 1 $\mu m$. The internal units for each fluid particle mass ($m_{f}$) and the time step ($\delta t$) were given the values of $10$ and $5\times10^{-4}$ respectively. Considering the blood density is equal to $1.025 g/cm^3$ \cite{kenner1989measurement} and assuming depth contribution to volume is equal to the unit, we can obtain the following equivalence between the simulation and the real system:

\begin{eqnarray}
\frac{m_f}{\mu m^{2}}  &=& \frac{1.025g}{cm^{2}}\label{eq:1}\\
  \nonumber m_f &=& \SI[per-mode=fraction,scientific-notation=true]{10.25}{\nano\gram}
\end{eqnarray}

The total mass of fluid within a channel with an area ($A_{f}$) equal to $540 \mu m^2$ was calculated to be $5.535 \mu g$. Due to the limited information available in the literature regarding the parasite's mass, we used cell density measurements from erythrocytes\cite{grover2011measuring}, human lung cancer cells, and mouse lymphoblastic leukemia cells \cite{bryan2014measuring} as a reference. Since these cell types have densities approximately equal to $1 g/cm^3$, we assumed the parasite's cell body density to be at least equal to that of water. Since the model is a two-dimensional representation, density is given as grams per unit of area, ($\rho_{ A_c}$): $1 g/cm^2$.

The scale units are the result of the relation between the physical system units (denoted with $'$) and the model units as follows:

\begin{eqnarray}
  M_{0} &=& \frac{m'}{m} \label{eq:3}\\
  T_{0} &=& \frac{t'}{\delta t} \label{eq:4}\\
  R_{0} &=& \frac{r'}{r}  \label{eq:5}\\
  PA_{0} &=& \frac{\rho A'}{\rho A} \label{eq:6}
\end{eqnarray}

Where $M_{0}$, $T_{0}$, $R_{0}$, and $PA_{0}$ are mass, time, length, and area density respectively.
With the preceding definition and taking into account that in the recording \cite{finkelsztein2015altering} the transition from discretization one to five takes $91$ [\emph{cs}], the equations from Equation \ref{eq:3} to Equation \ref{eq:6} were used to calculate the basic units scale factors between the physical system and the simulation:

\begin{eqnarray}
\nonumber M_{0} &=& \SI[{per-mode=fraction,scientific-notation=false}]{\fpeval{5.535/9000}}{\micro\gram} = \SI[scientific-notation=true]{0.615}{\nano\gram}\\
\nonumber T_{0} &=& \frac{0.91 s}{5 \times n_{disc.} \times n_{steps/disc.} \times n_{div. disc.} \times \delta t} = \frac{0.91 s}{50} = \SI[scientific-notation=true]{0.0182} {\second} \\
\nonumber R_{0} &=& \SI{1}{\micro\meter} \\ 
\nonumber PA_{0} &=& \SI[per-mode=fraction,scientific-notation=true]{\fpeval{1.025/16.666666666666668}}{\frac{\gram}{\centi\meter^2}} = \SI[scientific-notation=true]{0.06149999999999999}{\frac{\gram}{\centi\meter^2}}
\label{eq:8}
\end{eqnarray}

By using the previously obtained scale factors, velocity ($V_{0}$), acceleration ($a_{0}$) and energy ($E_0$) factors are derived:\\
\begin{eqnarray}
\nonumber V_{0} &=& \frac{R_{0}}{T_{0}} = \SI[{quotient-mode=fraction,scientific-notation=true}]{\fpeval{1/0.00182}}{\frac{\micro\meter}{\second}} = \SI[scientific-notation=true]{0.05494505494505495}{\frac{\centi\meter}{\second}}\\
\nonumber a_{0} &=& \frac{R_{0}}{T_{0}^{2}} = \SI[{quotient-mode=fraction,scientific-notation=true}]{\fpeval{1/0.033124}}{\frac{\micro\meter}{\second^{2}}} = \SI[scientific-notation=true]{30.18959062915107}{\frac{\centi\meter}{\second^{2}}}\\
\nonumber E_0 &=& \frac{M_0 R_0^2}{T_0^2} = \SI[scientific-notation=true]{1.8566e-18 }J \\
\label{eq:11}
\end{eqnarray}

\subsection{Parasite mass}
The parasite's mass is assumed to be uniformly distributed along its cellular body. The total mass is represented by the product of the number of nodes ($N_{n}$), and the mass of each node ($m_{n}$). This can also be expressed as the product of the parasite's area density ($\rho A_c$) and its area ($A_{Tc}$), as shown in Equation \ref{eq:9}:

\begin{equation}
  m_{n} N_{n}  = \rho A_c A_{Tc} \label{eq:9}
\end{equation}

The model uses 54 nodes to represent the parasite's structure, which has an estimated area ($A_{Tc}$) of approximately $32\mu m^{2}$. Based on these values, the mass of each node ($m_{n}$) can be calculated as follows:

\begin{eqnarray}
  \nonumber m_{n} &=& \frac{\rho A_c A_{Tc}}{N_{n}}\\
  \nonumber m_{n} &=& \frac{\SI[per-mode=fraction]{1}{\frac{\gram}{\meter^{-4}}}\cdot 32 m^{-12}}{54}\\
  \nonumber m_{n} &=& \SI[scientific-notation=true]{5.925925925925926}{\nano\gram}
\end{eqnarray}

\subsection{Reynolds number}

Reynolds number was obtained by following the approach proposed by Backer\cite{backer2005poiseuille} based on the velocity of the system:

\begin{equation}
  \langle V_{x} \rangle = \frac{\rho g_{x} D^{2}}{12\eta}
    \label{eq:12}
\end{equation}

Where $\rho$ is the system mass density which is assumed to remain constant during the simulation, $g_{x}$ is the external force applied to the particles, $\eta$ is the system dynamic viscosity, $D$ is half the length of the system box, and $V_{x}$ is the particles velocity measured parallel to $g_{x}$.

Given the expression Equation \ref{eq:10}:

\begin{equation}
  Re = \frac{\rho \langle V_{x}\rangle D}{\eta}\\
  \label{eq:10}
\end{equation}

Isolating $\eta$ from Equation \ref{eq:12} to replace it in Equation \ref{eq:10} allows calculating the Reynolds number in the simulation as:

\begin{equation}
  Re = \frac{12\langle V_{x}^{2}\rangle}{g_{x}D}\\
  \label{eq:13}
\end{equation}

Since we use internal units in the computational model, we need to convert between laboratory and simulation units. 
By using the scale factors previously calculated in Equation \ref{eq:8} we obtained that the scale factor to get the Reynolds number for the physical system is the unit:

\begin{eqnarray}
  Re_{0} &=& \frac{\left(\SI[{per-mode=fraction,scientific-notation=true}]{0.05494505494505495}{\frac{\centi\meter}{\second}}\right)^2}{\SI[{per-mode=fraction,scientific-notation=true}]{30.18959062915107}{\frac{\centi\meter}{\second^2}\cdot \SI[{per-mode=fraction,scientific-notation=true}]{0.0001}{\centi\meter}}}\\
  \nonumber Re_{0} &=& \SI[{per-mode=fraction,scientific-notation=true}]{1.0000000000000002}{}
\end{eqnarray}

\section{Numerical Experiments} \label{sec:numexp}
In order to understand whether our model presents some form of tropism or whether it can swim within a fluid, we ran a series of numerical experiments. Table \ref{tab:model_parameters} has the parameters used in them, unless otherwise stated. The data set has been made publicly available in Figshare, \cite{tcruzidataset}.

\begin{table}[H]
\centering
\caption{Model parameters description}
\begin{tabular}{llr}
\toprule
\textbf{Component} & \textbf{Parameter} & \textbf{Value} \\
\midrule
\textbf{Cell Body} & \texttt{kGlobal} (Spring constant) & 1 \\
 & \texttt{kAng} (Angle constraint coefficient) & 10 \\
 & \texttt{FedKd} (Area constraint coefficient for 2D systems) & 200 \\
 & \texttt{bdamp} (Damping coefficient) & 20 \\
 & \texttt{numberOfTransitionSTEPS} (Number of steps between discretizations) & 100 \\
\addlinespace
\textbf{Blood Flow} & \texttt{rcutdpd} (Cutoff distance, defines whether particles interact or not) & 1 \\
 & \texttt{DPDAcc} (External force applied to implement Poiseuille flow) & 0.055 \\
 & \texttt{sigma} (Random force noise amplitude) & 4.5 \\
 & \texttt{gamma} (Strength of dissipative forces) & 20 \\
 & \texttt{goalVel} (Target Re velocity) & 1.25 \\
\addlinespace
\textbf{Coupled Simulation} & \texttt{kTcruzi} (Structure-DPD interaction coefficient) & 200 \\
\bottomrule
\end{tabular}
\label{tab:model_parameters}
\end{table}

\begin{figure}[H]
  \centering
  \includegraphics[width=0.9\columnwidth]{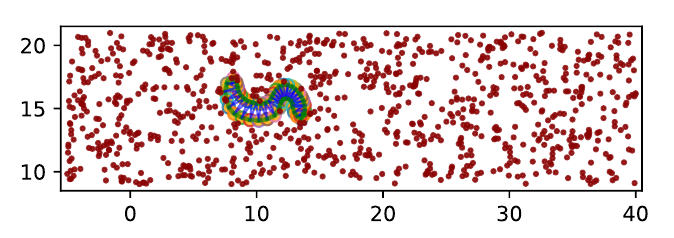}
  \caption{T. Cruzi 2D model: thin straight lines represent internal springs (\emph{blue}), thick straight lines the surface of the parasite (\emph{green}), and thick disks the nodes (\emph{yellow}). Thin disks represent the DPD particles (\emph{red}). In this case the simulation box has width of  $45 \mu m$ and a depth of 12 $\mu m$.} 
  \label{fig:snapDens1}
\end{figure}

An example of the simulation box and the T. Cruzi model within a fluid of DPD particles is depicted in Figure \ref{fig:snapDens1}. The axis coordinates correspond to the laboratory coordinates, in $\mu m$.

The parasite model is placed within the simulation box, surrounded by DPD particles moving from right to left. Its shape evolves periodically within the five discretizations as described before. Typically, the parasite model end to end distance varies from 3 $\mu m$ up to 8 $\mu m$. 


\subsection{No flow displacement} 
\begin{figure}[H]
  \centering
  \includegraphics[width=0.9\columnwidth]{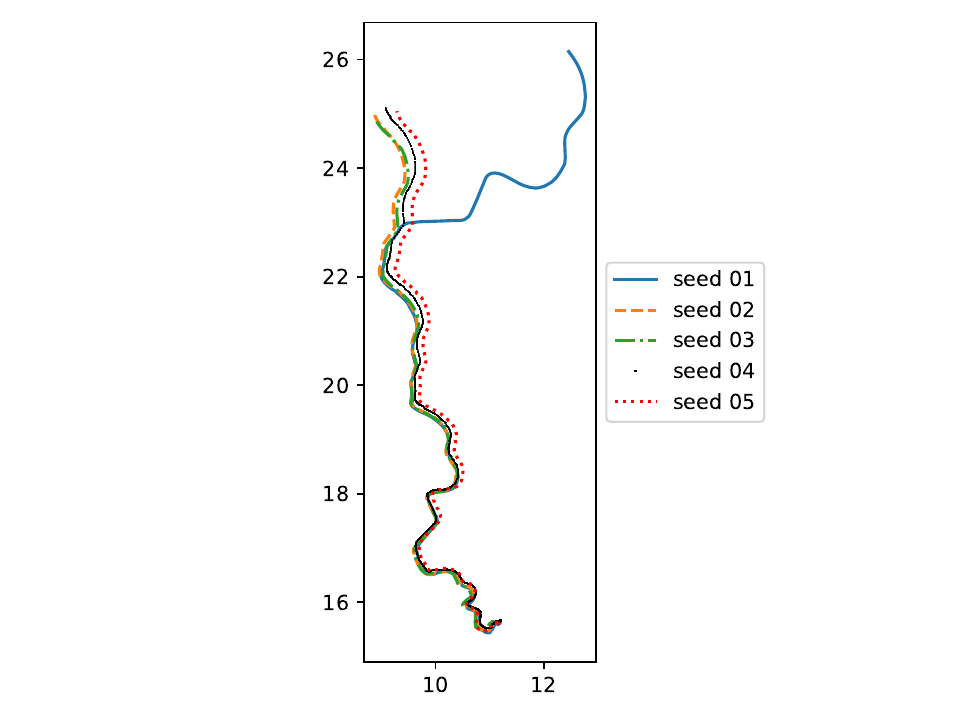}
  \caption{Center of mass position of the T.Cruzi as a function of time for different values of the random number generator seed: thick continuous  \emph{(blue)}, first seed; dashed \emph{(yellow)}, second seed; dash dot \emph{(green)}, third seed; thin continuous \emph{(black)}, fourth seed; dotted  \emph{(red)}, fifth seed.(\emph{Color online}). Distances in $\mu m$.}
  \label{fig:diferentes_trayectorias_002_Alberto_TCruzi_recording_08_nueva_900_particulas_diferentes_semillas.pdf}
\end{figure}

In our initial set of experiments, we investigated whether the deformation of the parasite extracted from an experimental setup \cite{finkelsztein2015altering} induces a directed displacement within a stationary fluid.
Figure \ref{fig:diferentes_trayectorias_002_Alberto_TCruzi_recording_08_nueva_900_particulas_diferentes_semillas.pdf} illustrates the trajectory of the parasite's center of mass for different values of the random number generator seed. The initial coordinate of the center of mass is \((11.15, 15.62)\). The five trajectories follow a similar path, mostly in the vertical direction. This implies that the trajectory depends on the interaction between the fluid and the parasite model; and the specific route does not depend on the random fluctuations. 

\begin{figure}[H]
  \centering
  \includegraphics[width=0.9\columnwidth]{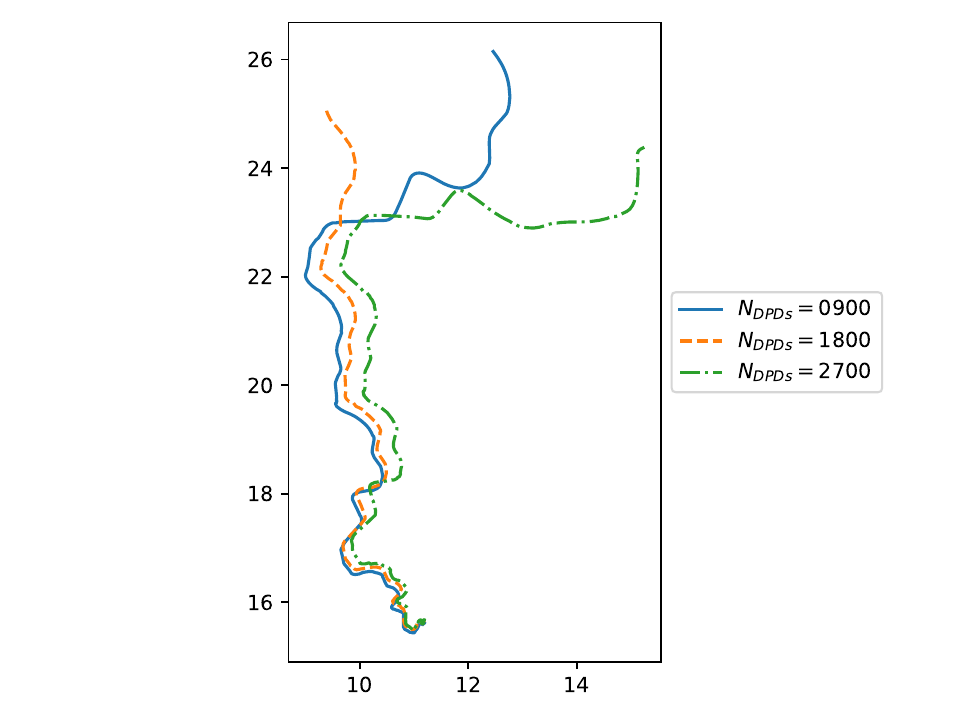}
  \caption{Center of mass position of the T. Cruzi as a function of time for different DPD particle densities: thick smooth continuous \emph{(blue)}, \(N_{DPD}=900\); dashed  \emph{(yellow)}, \(N_{DPD}=1800\); dot dashed \emph{(green)}, \(N_{DPD}=2700\).(\emph{Color online}). Distances in $\mu m$.}
  \label{fig:diferentes_trayectorias_002_Alberto_TCruzi_recording_10_nueva_cambio_numero_de_particulas.pdf}
\end{figure}

Figure \ref{fig:diferentes_trayectorias_002_Alberto_TCruzi_recording_10_nueva_cambio_numero_de_particulas.pdf} compares the displacement of the center of mass for simulations using 900, 1800, and 2700 DPD particles. The resulting paths followed by the parasite are nearly identical, indicating that particular trajectories depend on the interaction between the model and the fluid.  


\begin{figure}[H]
\centering
\includegraphics[width=0.9 \columnwidth]{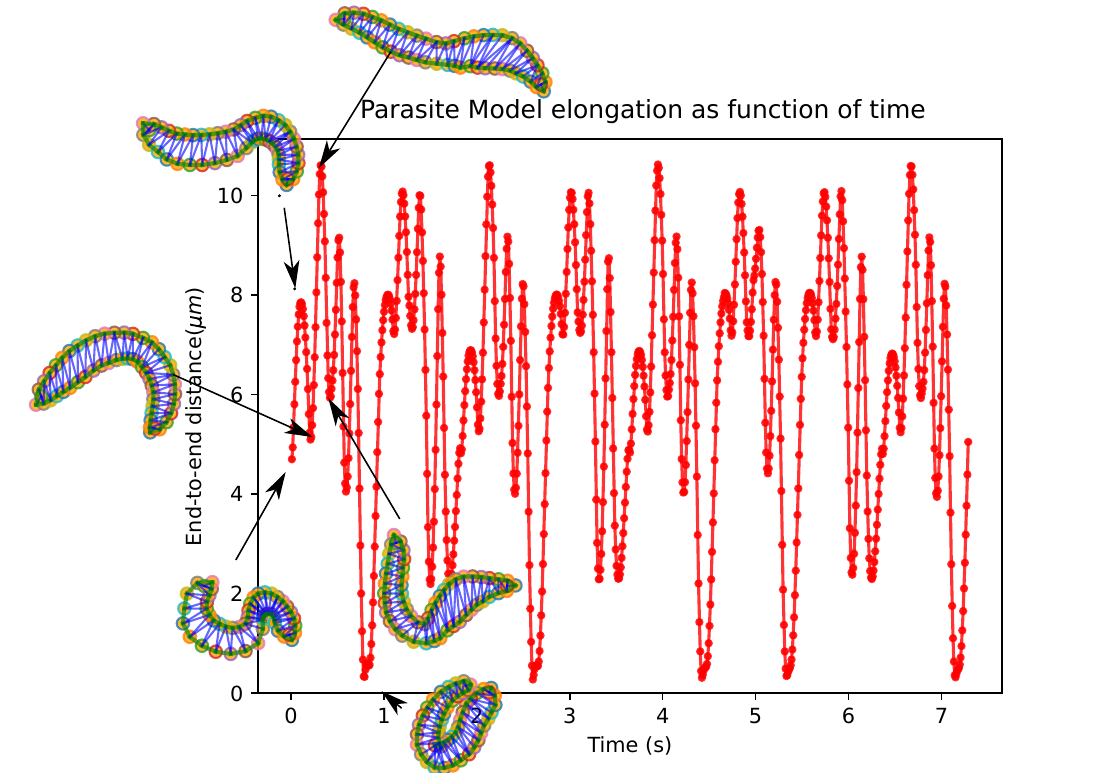}
\caption{Elongation of the T. Cruzi model as a function of time as it changes within the five discretizations. Unit: $\mu m$. End-to-end distance varies between $\SI[round-precision=0]{1}{\micro\meter}$ and $\SI[round-precision=1]{10.5}{\micro\meter}$. The image shows a cyclic behavior resulting from the transition between discretizations. On top of the figure we have included different geometries of the parasite, and arrows pointing to the point in the figure that they represent.}
\label{fig:07ElnFTime_08_nueva_900_particulas_diferentes_semillas_01_seed.pdf}
\end{figure}

As shown in Figure \ref{fig:diferentes_trayectorias_002_Alberto_TCruzi_recording_10_nueva_cambio_numero_de_particulas.pdf}, the model undergoes a displacement of about 10 $\mu m$ in the vertical direction, irrespective of the number of DPD particles or the random number generator seed. The end-to-end length of the parasite model ranges between 1 and 10 $\mu m$, as seen in Figure \ref{fig:07ElnFTime_08_nueva_900_particulas_diferentes_semillas_01_seed.pdf}. The displacement observed in the simulations is significant because its magnitude is of the same order of magnitude as the parasite's end-to-end body length.


Since both the magnitude of the displacement and the general shape of the center of mass trajectory do not depend on the DPD number densities, we conclude that the model is capable of self-propulsion in a still fluid. 


\begin{figure}[H]
  \centering
  \includegraphics[width=0.9\columnwidth]{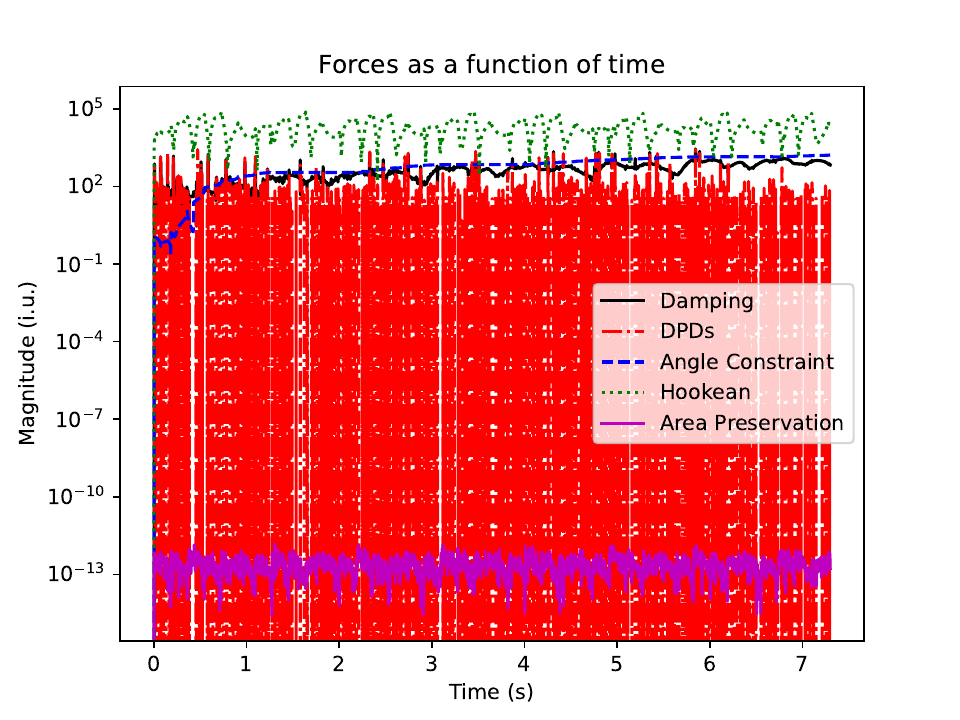}
  \caption{Force magnitude as a function of time: dotted (\emph{green}) line, springs Hookean; continuous (\emph{black}), damping; dashed
    (\emph{blue}), angle constrain; dot dashed (\emph{red}), DPD forces over the structure.}
  \label{fig:13_logeveryForceEvolution.pdf}
\end{figure}

We present the magnitude of the different forces that act on the parasite as a function of time in Figure \ref{fig:13_logeveryForceEvolution.pdf}. The faintest forces are those that exert the angle constraint. In magnitude this force is also quite stable. There is a correlation between the damping forces and the forces exerted over the parasite by the DPD particles. The largest contribution comes out of the Hookean forces, which on average are two orders of magnitude stronger than the other forces on this system.

\subsection{Laminar flow - without internal deformation}
With these experiments we want to explore whether the periodic deformation of the model parasite has an effect on the directed movement within the fluid -which results from the DPD particles moving from right to left. In other words, whether periodic movement plus fluid flow leads to a form of swimming.

The heart, esophagus, and colon, being the most relevant parasitized organs in the human body during chronic stage, were selected as the focus for configuring the fluid model. We referred to the values reported in the literature for the Reynolds numbers of blood flow, in order to identify the values that needed to be used in our simulation. The Table \ref{ReyOrgans} presents the values and references of those values. It is important to recognize that these values correspond to the fluid flow, and are not to be confused with the average speed of the parasite in stationary fluid, reported by \cite{arias2020motility}.

\begin{table}[H]
\begin{center}
\begin{tabular}{ | c | c | c | c |}
\hline
\textbf{Organ} & \textbf{Artery} & \textbf{Blood Reynolds number} & \textbf{Reference}\\ \hline
Colon & Superior and inferior mesenteric & 702–1424 & \cite{lee2002numerical} \\ \hline
Heart & Right coronary & 150 & \cite{wootton1999fluid} \\ \hline
Esophagus (Thoracic) & Thyroid (Aorta communication) & 600 & \cite{ku1997blood}\\ \hline
\end{tabular}
\end{center}
 \caption{Reynolds numbers for fluid flow simulation in different arteries}
\label{ReyOrgans}
\end{table}

We replicated each experiment three times per organ, using different seeds to initialize the particle positions. We explored the response of the \emph{T cruzi} to different flow conditions in terms of energy, displacement, and cell deformation.

\begin{figure}[H]
  \centering
  \includegraphics[width=0.6\columnwidth]{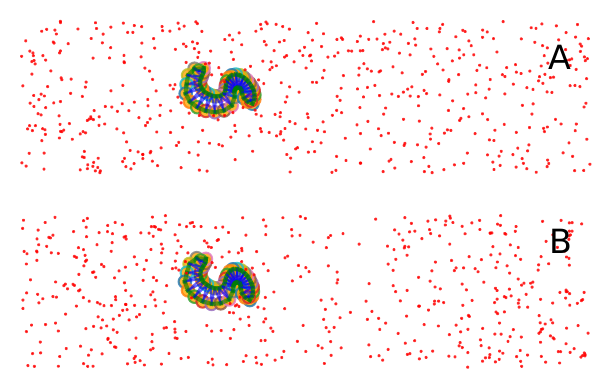}
  \caption{Baseline structure with no motion induced. A) The initial position of the structure is presented, located horizontally in the middle of the channel. B) After the simulation, the structure still lies in the same location.}
  \label{fig:starting_ending}
\end{figure}

To establish a baseline, we decided to first explore the effect of the fluid over the parasite model structure which does not deform nor is subject to external fluid motion. This experiment would not be feasible with the real parasite, thus the comparison is only possible with a computational model. As expected, there is no displacement in the parasite model, as can be seen in Figure \ref{fig:starting_ending}. We should remind the readers that, at this scale, although gravity is still present. However, its direction would be ``into the paper'', and therefore does not need to be included in this simulation.


The following numerical experiment consists in subjecting the parasite model, still without including internal deformation, to the external fluid motion, mass under different Reynolds numbers. 

\begin{figure}[H]
  \centering
  \includegraphics[width=0.9\columnwidth]{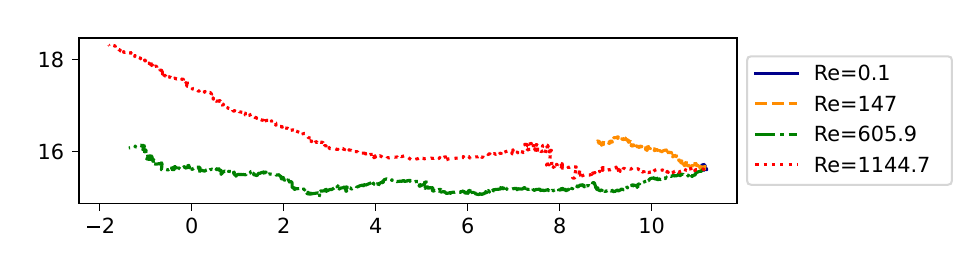}
  \caption{Center of mass displacement of the parasite model without internal deformation, for different values of the Reynolds Number. Results for different Reynolds numbers are presented: Continuous line \emph{(blue)}, $0.1$ \emph{Re}; dashed \emph{(orange)}, $147$\emph{Re}; dash dot  \emph{(green)}, $605.9$\emph{Re}; dotted \emph{(red)}, $1144.7$\emph{Re}. }
  \label{fig:polygon_mass_center}
\end{figure}

In Figure \ref{fig:polygon_mass_center} we observe the effect of the Reynolds number on the structure's trajectory. At a very low Reynolds number $(Re = 0.1)$, the structure's movement is minimal, and it remains near its initial position. As the Reynolds number increases, the external flow has a more pronounced effect, dragging the structure downstream. 

\subsection{Laminar flow - with internal deformation}
In the following sections we refer to the parasite model with internal deformation.

\subsubsection{Energies}
\begin{figure}[H]
\centering
\begin{subfigure}[c]{0.9\columnwidth}
\includegraphics[width=0.9\columnwidth]{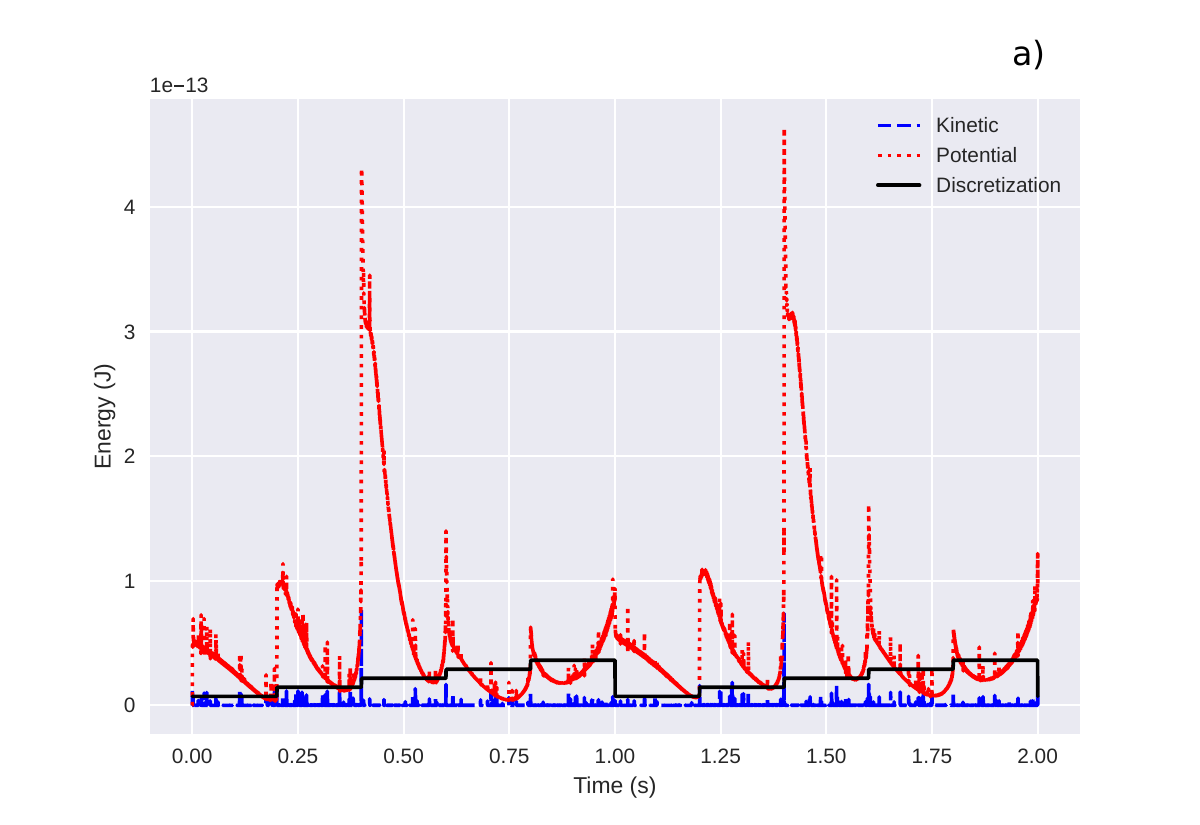}
\caption{}
\label{fig:13a}
\end{subfigure}\hfill

\begin{subfigure}[c]{0.9\columnwidth}
\includegraphics[width=0.9\columnwidth]{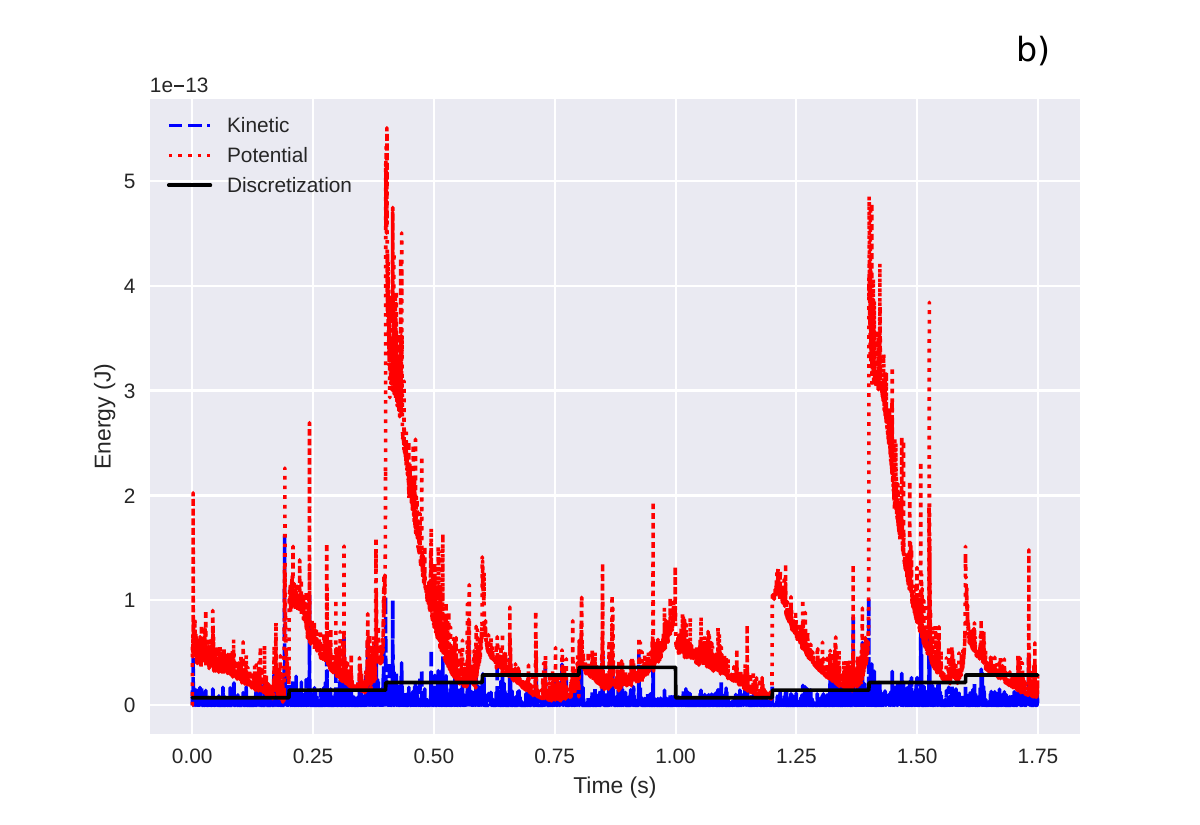}
\caption{}
\label{fig:13b}
\end{subfigure}

\caption{Evolution of kinetic and potential energy: dash dot \emph{(blue)}, kinetic energy; smooth continuous \emph{(black)}, discretization transition; dotted \emph{(red)}, potential energy. \ref{fig:13a} Simulation in a stationary fluid. \ref{fig:13b} Simulation with flow at Reynolds = 1147.74. The random number generator seed value is 1.}
\label{fig:08Energias_Horizontal_ComparedSeed1399.png}
\end{figure}

The energy dynamics of the model provide insight into the biomechanical cost of locomotion. Figure \ref{fig:08Energias_Horizontal_ComparedSeed1399.png} shows the energy profile of the model under two different flow conditions.


In \ref{fig:13a} we focus on measuring the energy demanded by the structure to produce what we have defined as its natural motion, it is, self-propulsion or swimming in the absence of flow. The amount of energy stored by the structure increases in response to the induced migration between discretizations as described in section \ref{sec:parasite-fluid-interaction}. The forces exerted during the migration process result in a sharp increase in the stored potential energy, which peaks at approximately $\SI[round-precision=0]{4e-13}{Joules}$ before being dissipated as the structure reaches the next stable state.

The introduction of an external flow alters the model's energy dynamics \ref{fig:13b}. The primary effect of the flow is a significant increase in the parasite's kinetic energy, which fluctuates around $\SI[round-precision=1]{0.5e-13}{Joules}$. Additionally, the potential energy peak increases to approximately $\SI[round-precision=0]{7e-13}{Joules}$. \\


\textbf{Potential} \\

To investigate the influence of the Reynolds number on the parasite, we analyzed the potential energy distributions from simulations at four different Re values.

\begin{figure}[H]
  \centering
  \includegraphics[width=0.9\columnwidth]{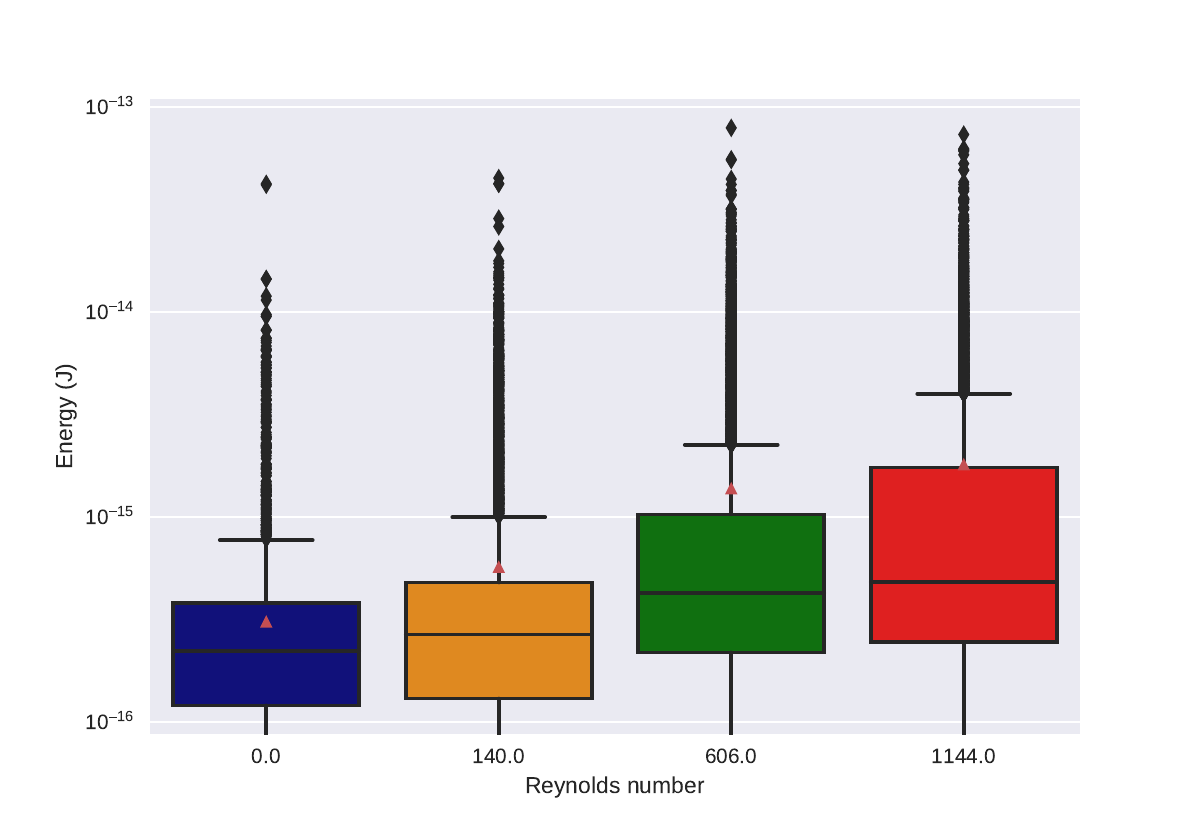}
  \caption{Parasite potential energy comparison at different Reynolds numbers: $0.1$\emph{Re} \emph{(blue)}; $147$\emph{Re} \emph{(orange)}; $605.93$\emph{Re} \emph{(green)}; $1144.77$\emph{Re}\emph{(red)}. }
  \label{fig:13EnergiaPotencial_Horizontal_ComparedSeed2.png}
\end{figure}

As shown in Figure \ref{fig:13EnergiaPotencial_Horizontal_ComparedSeed2.png}, the potential energy distributions are qualitatively similar across the tested range of Reynolds numbers. Although a slight upward trend in the median energy is observable with increasing Re, the interquartile ranges are comparable and show significant overlap. This suggests that while a relationship between the Reynolds number and the potential energy may exist, it is rather weak. \\



\begin{figure}[H]
  \centering
  \includegraphics[width=0.9\columnwidth]{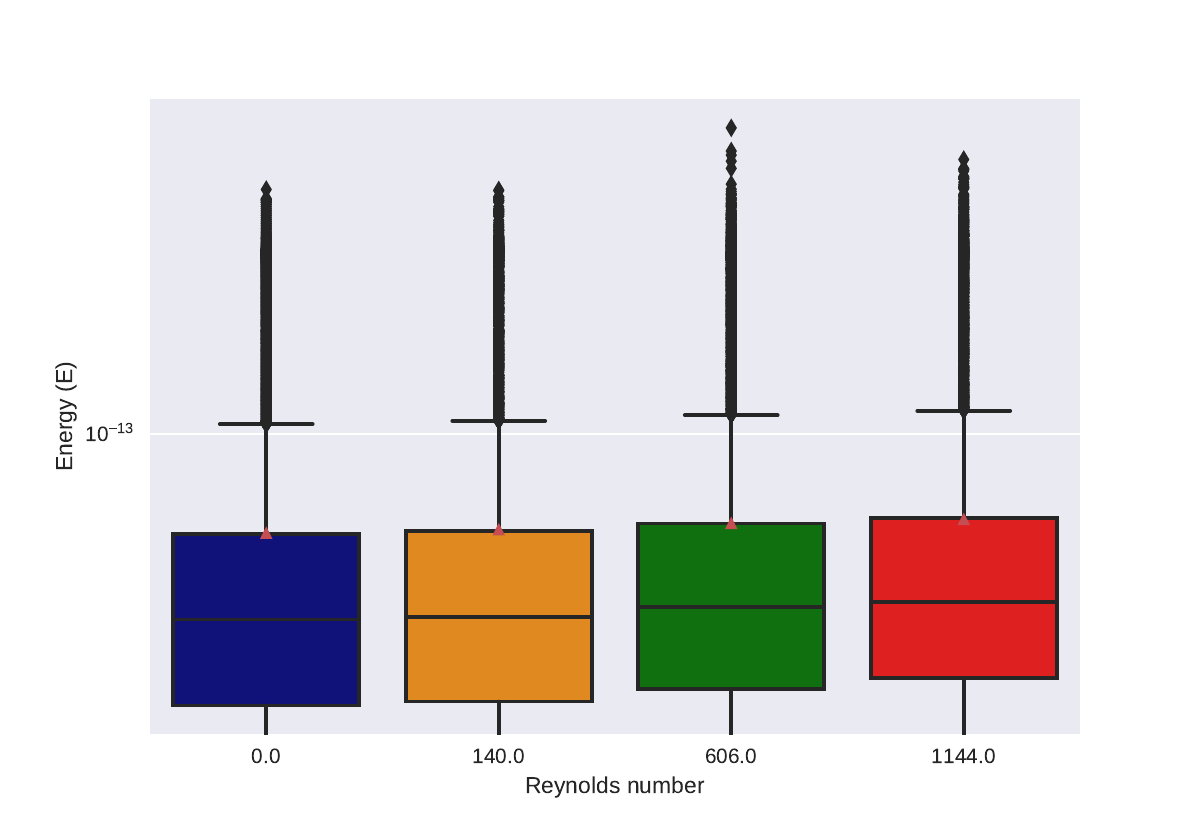}
  \caption{Parasite kinetic energy comparison at different Reynolds numbers: $0.22$\emph{Re} \emph{(blue)}; $142.32$\emph{Re} \emph{(orange)}; $608.03$\emph{Re} \emph{(green)}; $1145.82$\emph{Re}\emph{(red)}. }
  \label{fig:12EnergiaCinetica_Horizontal_ComparedSeed3.png}
\end{figure}

\textbf{Kinetic}\\
In \ref{fig:12EnergiaCinetica_Horizontal_ComparedSeed3.png} we found evidence of flow-induced variations in the kinetic energy of the parasite. The magnitude of kinetic energy and the Reynolds numbers are clearly correlated.

As can be seen in Figure \ref{fig:12EnergiaCinetica_Horizontal_ComparedSeed3.png}, there are no significant differences among the four different distribution of kinetic energy values. Nor in the skewness of the distributions nor in the mean or median.  

In light of our results, parasite motility is affected by the flow because the energy required for motility increases with the flow velocity. Given this increased energy demand, high Reynolds number flows can be considered hostile to the main processes that enable parasite proliferation: motion and cell invasion.

\subsubsection{Displacement}

Our experiments suggest that at low Reynolds numbers, the parasite's own motility could be more effective, whereas at high Reynolds numbers, its trajectory is likely dominated by the fluid drag driving the cell upwards. This is consistent with the findings from \cite{uppaluri2011unicellular} where the parasite's displacement was faster and upstream, following a wavy trajectory.


\begin{figure}[H]
  \centering
  \includegraphics[width=0.9\columnwidth]{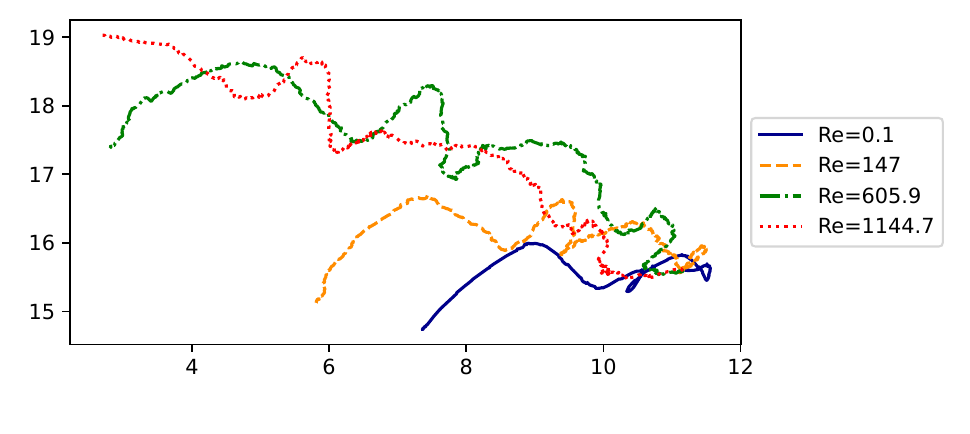}
  \caption{Center of mass displacement of the parasite model under laminar flow towards the left: smooth continuous \emph{(blue)}, $0.1$\emph{Re};  dashed \emph{(orange)}, $147$\emph{Re}; dash dot \emph{(green)}, $605.9$\emph{Re}; dotted \emph{(red)}, $1144.7$\emph{Re}. The coordinates of the initial position of the structure is the point $(11.14, 15.61)$. Distances in $\mu m$} 
  \label{fig:17CenterOfMass2D_Horizontal_ComparedSeed1.png}
\end{figure}


The results presented in \ref{fig:17CenterOfMass2D_Horizontal_ComparedSeed1.png} could explain, to some extent, how the parasite travels within the human body, reaching organs far from the entry wound: taking advantage of the drag produced by the flow. While it may seem obvious that a viscous fluid would drag a small body along, the model allows us to relate the drag to the intrinsic movement of the parasite and, in turn, explore the role of tropism. It also seems possible that, for low Reynolds numbers, chemotaxis-driven motility could be facilitated.

\subsubsection{Cell deformation}
Cell deformation is a process that can be either reversible or permanent and several factors like cell viscoelasticity, viscoplasticity, cell geometry or the time scale during which the stimulus is applied play an important role in the outcome \cite{molnar2021plastic}. Our representation of the cell body presents minor changes on its end-to-end distance \ref{fig:07ElnFTime_Horizontal_ComparedSeed2.png} when exposed to different Reynolds numbers, and in all cases those changes were reversible: 

\begin{figure}[H]
  \centering
  \includegraphics[width=0.9\columnwidth]{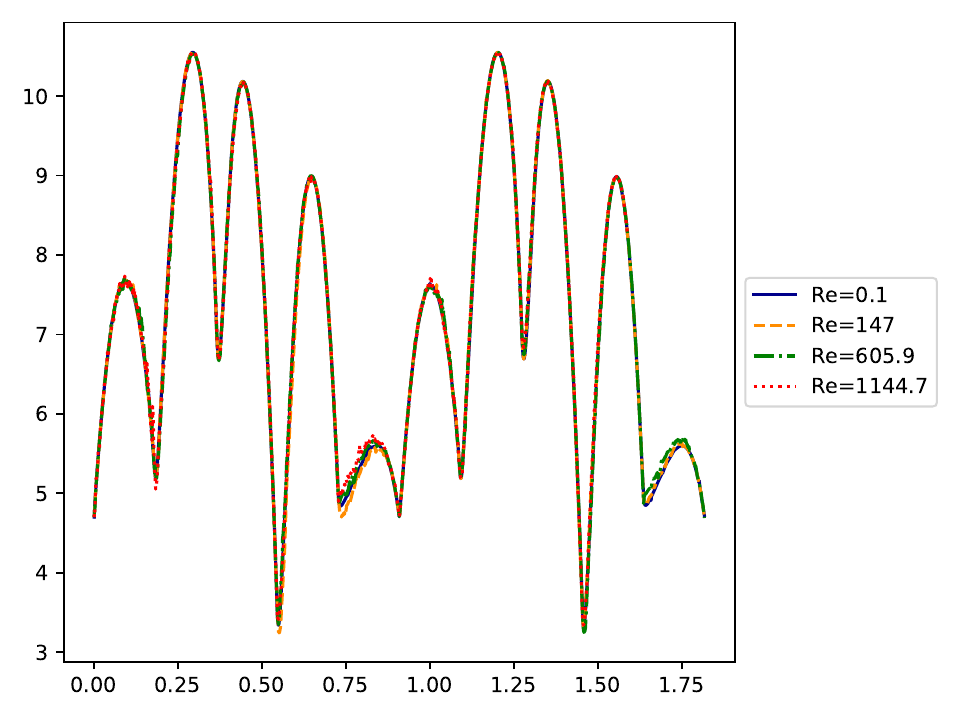}  
  \caption{End-to-end distance measured for different Reynolds numbers: smooth continuous \emph{(blue)}, $0.1$\emph{Re};  dashed \emph{(light blue)}, $147$\emph{Re}; dash dot \emph{(brown)}, $605.93$\emph{Re}; dotted \emph{(yellow)}, $1144.77$\emph{Re}. Distances in $\mu m$}
  \label{fig:07ElnFTime_Horizontal_ComparedSeed2.png}
\end{figure}

The end-to-end distance exhibits consistent behavior across all cases \ref{fig:07ElnFTime_Horizontal_ComparedSeed2.png}. Although minor deviations are observed at the highest Re, these are negligible. The overall result indicates that the cell's elongation is largely independent of the external flow velocity. This finding is consistent with \cite{uppaluri2011unicellular} where experiments with live cells demonstrated that cell elongation has little dependence on the surrounding flow velocity.

\section{Conclusions and closing remarks} \label{sec:conclusion}

We have introduced a 2D model for the \emph{Trypanosoma Cruzi} parasite and have characterized its interaction with a 2D fluid of DPD particles. In a first set of experiments we studied the motion of the center of mass for different values of the density of DPD particles, in the absence of forced external particle flow. We found no evidence of significant changes in the path followed by the center of mass of the model parasite as a function of the particle density, nor of the seed of the random number generator. 

The displacement of the center of mass is also relatively large compared to the deformation of the body of the model parasite, the former being of the order of 10 $\mu m$, the latter varying between 1 and 10 $\mu m$. In the case of the laminar flow, the experiments show how the parasite model moves in the direction of the flow; and the displacement is stronger for larger Reynolds numbers.




Therefore, the results from both the no-flow and laminar flow simulations indicate that our coupled model does show evidence of a relationship between that cell deformation and the direction and magnitude of the parasite's displacement. In plain words, the parasite model presented in this paper is able to self-propel in a stationary fluid but is passively transported when subjected to an external flow.

Our model's ability to self-propel in a stationary fluid is consistent with experimental observations \cite{arias2020motility}, which supports its biological relevance. However, our results further suggest that this autonomous motility is insufficient for long-range transport. We therefore conclude that the long-distance displacement of the parasite within a host is primarily caused by the blood flow itself, rather than by the parasite's own propulsion.


We acknowledge the fact that our model is a very simplified version of the reality. The most plausible improvements that need to be introduced in the model are: a 3D modeling of the parasite, a more realistic model of the blood flow, and the inclusion of the flagellum. 


In our concept, the blood flow plays a key role on cellular tropism, either because it drags the trypomastigotes to environments where more resources are available or because it allows the parasite to drive its motion by the chemical signal of potential host cells. 

We found no experiments that could confirm the order of magnitude of the forces or other parameters for the simulation. Nonetheless, we set up the simulations to mimic the movement shown in videos of the parasites. That is, we also found some motility patterns similar to those reported from \emph{in vivo} experiments.

However, confirmation of these results in terms of a real environment makes it necessary to develop a 3D model of the parasite, as well as its interaction with a flow formed by 3D red blood cells.

\section*{Competing interests}

The authors declare no competing interests, either financial nor
non-financial that are directly or indirectly related to this article.

\section*{Bibliography}
\bibliography{dpd}

\begin{thebibliography}{21}%
\makeatletter
\providecommand \@ifxundefined [1]{%
 \@ifx{#1\undefined}
}%
\providecommand \@ifnum [1]{%
 \ifnum #1\expandafter \@firstoftwo
 \else \expandafter \@secondoftwo
 \fi
}%
\providecommand \@ifx [1]{%
 \ifx #1\expandafter \@firstoftwo
 \else \expandafter \@secondoftwo
 \fi
}%
\providecommand \natexlab [1]{#1}%
\providecommand \enquote  [1]{``#1''}%
\providecommand \bibnamefont  [1]{#1}%
\providecommand \bibfnamefont [1]{#1}%
\providecommand \citenamefont [1]{#1}%
\providecommand \href@noop [0]{\@secondoftwo}%
\providecommand \href [0]{\begingroup \@sanitize@url \@href}%
\providecommand \@href[1]{\@@startlink{#1}\@@href}%
\providecommand \@@href[1]{\endgroup#1\@@endlink}%
\providecommand \@sanitize@url [0]{\catcode `\\12\catcode `\$12\catcode
  `\&12\catcode `\#12\catcode `\^12\catcode `\_12\catcode `\%12\relax}%
\providecommand \@@startlink[1]{}%
\providecommand \@@endlink[0]{}%
\providecommand \url  [0]{\begingroup\@sanitize@url \@url }%
\providecommand \@url [1]{\endgroup\@href {#1}{\urlprefix }}%
\providecommand \urlprefix  [0]{URL }%
\providecommand \Eprint [0]{\href }%
\providecommand \doibase [0]{https://doi.org/}%
\providecommand \selectlanguage [0]{\@gobble}%
\providecommand \bibinfo  [0]{\@secondoftwo}%
\providecommand \bibfield  [0]{\@secondoftwo}%
\providecommand \translation [1]{[#1]}%
\providecommand \BibitemOpen [0]{}%
\providecommand \bibitemStop [0]{}%
\providecommand \bibitemNoStop [0]{.\EOS\space}%
\providecommand \EOS [0]{\spacefactor3000\relax}%
\providecommand \BibitemShut  [1]{\csname bibitem#1\endcsname}%
\let\auto@bib@innerbib\@empty
\bibitem [{\citenamefont {Villalta}\ \emph {et~al.}(2009)\citenamefont
  {Villalta}, \citenamefont {Scharfstein}, \citenamefont {Ashton},
  \citenamefont {Tyler}, \citenamefont {Guan}, \citenamefont {Mukherjee},
  \citenamefont {Lima}, \citenamefont {Alvarez}, \citenamefont {Weiss},
  \citenamefont {Huang} \emph {et~al.}}]{villalta2009perspectives}%
  \BibitemOpen
  \bibfield  {author} {\bibinfo {author} {\bibfnamefont {F.}~\bibnamefont
  {Villalta}}, \bibinfo {author} {\bibfnamefont {J.}~\bibnamefont
  {Scharfstein}}, \bibinfo {author} {\bibfnamefont {A.~W.}\ \bibnamefont
  {Ashton}}, \bibinfo {author} {\bibfnamefont {K.~M.}\ \bibnamefont {Tyler}},
  \bibinfo {author} {\bibfnamefont {F.}~\bibnamefont {Guan}}, \bibinfo {author}
  {\bibfnamefont {S.}~\bibnamefont {Mukherjee}}, \bibinfo {author}
  {\bibfnamefont {M.~F.}\ \bibnamefont {Lima}}, \bibinfo {author}
  {\bibfnamefont {S.}~\bibnamefont {Alvarez}}, \bibinfo {author} {\bibfnamefont
  {L.~M.}\ \bibnamefont {Weiss}}, \bibinfo {author} {\bibfnamefont
  {H.}~\bibnamefont {Huang}}, \emph {et~al.},\ }\bibfield  {title} {\bibinfo
  {title} {Perspectives on the trypanosoma cruzi--host cell receptor
  interactions},\ }\href@noop {} {\bibfield  {journal} {\bibinfo  {journal}
  {Parasitology research}\ }\textbf {\bibinfo {volume} {104}},\ \bibinfo
  {pages} {1251} (\bibinfo {year} {2009})}\BibitemShut {NoStop}%
\bibitem [{\citenamefont {Stanaway}\ and\ \citenamefont
  {Roth}(2015)}]{stanaway2015burden}%
  \BibitemOpen
  \bibfield  {author} {\bibinfo {author} {\bibfnamefont {J.~D.}\ \bibnamefont
  {Stanaway}}\ and\ \bibinfo {author} {\bibfnamefont {G.}~\bibnamefont
  {Roth}},\ }\bibfield  {title} {\bibinfo {title} {The burden of chagas
  disease: estimates and challenges},\ }\href@noop {} {\bibfield  {journal}
  {\bibinfo  {journal} {Global Heart}\ }\textbf {\bibinfo {volume} {10}},\
  \bibinfo {pages} {139} (\bibinfo {year} {2015})}\BibitemShut {NoStop}%
\bibitem [{\citenamefont {Organization}\ \emph {et~al.}(2012)\citenamefont
  {Organization} \emph {et~al.}}]{world2012research}%
  \BibitemOpen
  \bibfield  {author} {\bibinfo {author} {\bibfnamefont {W.~H.}\ \bibnamefont
  {Organization}} \emph {et~al.},\ }\bibfield  {title} {\bibinfo {title}
  {Research priorities for chagas disease, human african trypanosomiasis and
  leishmaniasis.},\ }\href@noop {} {\bibfield  {journal} {\bibinfo  {journal}
  {World Health Organization technical report series}\ ,\ \bibinfo {pages} {v}}
  (\bibinfo {year} {2012})}\BibitemShut {NoStop}%
\bibitem [{\citenamefont {Arias-del Angel}\ \emph {et~al.}(2020)\citenamefont
  {Arias-del Angel}, \citenamefont {Santana-Solano}, \citenamefont
  {Santill{\'a}n},\ and\ \citenamefont {Manning-Cela}}]{arias2020motility}%
  \BibitemOpen
  \bibfield  {author} {\bibinfo {author} {\bibfnamefont {J.~A.}\ \bibnamefont
  {Arias-del Angel}}, \bibinfo {author} {\bibfnamefont {J.}~\bibnamefont
  {Santana-Solano}}, \bibinfo {author} {\bibfnamefont {M.}~\bibnamefont
  {Santill{\'a}n}},\ and\ \bibinfo {author} {\bibfnamefont {R.~G.}\
  \bibnamefont {Manning-Cela}},\ }\bibfield  {title} {\bibinfo {title}
  {Motility patterns of trypanosoma cruzi trypomastigotes correlate with the
  efficiency of parasite invasion in vitro},\ }\href@noop {} {\bibfield
  {journal} {\bibinfo  {journal} {Scientific Reports}\ }\textbf {\bibinfo
  {volume} {10}},\ \bibinfo {pages} {15894} (\bibinfo {year}
  {2020})}\BibitemShut {NoStop}%
\bibitem [{\citenamefont {Ballesteros-Rodea}\ \emph {et~al.}(2012)\citenamefont
  {Ballesteros-Rodea}, \citenamefont {Santillán}, \citenamefont
  {Martínez-Calvillo},\ and\ \citenamefont {Manning-Cela}}]{Ballesteros12}%
  \BibitemOpen
  \bibfield  {author} {\bibinfo {author} {\bibfnamefont {G.}~\bibnamefont
  {Ballesteros-Rodea}}, \bibinfo {author} {\bibfnamefont {M.}~\bibnamefont
  {Santillán}}, \bibinfo {author} {\bibfnamefont {S.}~\bibnamefont
  {Martínez-Calvillo}},\ and\ \bibinfo {author} {\bibfnamefont
  {R.}~\bibnamefont {Manning-Cela}},\ }\bibfield  {title} {\bibinfo {title}
  {Flagellar motility of trypanosoma cruzi epimastigotes},\ }\href
  {https://doi.org/https://doi.org/10.1155/2012/520380} {\bibfield  {journal}
  {\bibinfo  {journal} {BioMed Research International}\ }\textbf {\bibinfo
  {volume} {2012}},\ \bibinfo {pages} {520380} (\bibinfo {year} {2012})},\
  \Eprint
  {https://arxiv.org/abs/https://onlinelibrary.wiley.com/doi/pdf/10.1155/2012/520380}
  {https://onlinelibrary.wiley.com/doi/pdf/10.1155/2012/520380} \BibitemShut
  {NoStop}%
\bibitem [{\citenamefont {Finkelsztein}\ \emph {et~al.}(2015)\citenamefont
  {Finkelsztein}, \citenamefont {Diaz-Soto}, \citenamefont {Vargas-Zambrano},
  \citenamefont {Suesca}, \citenamefont {Guzm{\'a}n}, \citenamefont
  {L{\'o}pez}, \citenamefont {Thomas}, \citenamefont {Forero-Shelton},
  \citenamefont {Cuellar}, \citenamefont {Puerta} \emph
  {et~al.}}]{finkelsztein2015altering}%
  \BibitemOpen
  \bibfield  {author} {\bibinfo {author} {\bibfnamefont {E.~J.}\ \bibnamefont
  {Finkelsztein}}, \bibinfo {author} {\bibfnamefont {J.~C.}\ \bibnamefont
  {Diaz-Soto}}, \bibinfo {author} {\bibfnamefont {J.~C.}\ \bibnamefont
  {Vargas-Zambrano}}, \bibinfo {author} {\bibfnamefont {E.}~\bibnamefont
  {Suesca}}, \bibinfo {author} {\bibfnamefont {F.}~\bibnamefont {Guzm{\'a}n}},
  \bibinfo {author} {\bibfnamefont {M.~C.}\ \bibnamefont {L{\'o}pez}}, \bibinfo
  {author} {\bibfnamefont {M.~C.}\ \bibnamefont {Thomas}}, \bibinfo {author}
  {\bibfnamefont {M.}~\bibnamefont {Forero-Shelton}}, \bibinfo {author}
  {\bibfnamefont {A.}~\bibnamefont {Cuellar}}, \bibinfo {author} {\bibfnamefont
  {C.~J.}\ \bibnamefont {Puerta}}, \emph {et~al.},\ }\bibfield  {title}
  {\bibinfo {title} {Altering the motility of trypanosoma cruzi with rabbit
  polyclonal anti-peptide antibodies reduces infection to susceptible mammalian
  cells},\ }\href@noop {} {\bibfield  {journal} {\bibinfo  {journal}
  {Experimental parasitology}\ }\textbf {\bibinfo {volume} {150}},\ \bibinfo
  {pages} {36} (\bibinfo {year} {2015})}\BibitemShut {NoStop}%
\bibitem [{\citenamefont {Hoogerbrugge}\ and\ \citenamefont
  {Koelman}(1992)}]{Hoogerbrugge_1992}%
  \BibitemOpen
  \bibfield  {author} {\bibinfo {author} {\bibfnamefont {P.~J.}\ \bibnamefont
  {Hoogerbrugge}}\ and\ \bibinfo {author} {\bibfnamefont {J.~M. V.~A.}\
  \bibnamefont {Koelman}},\ }\bibfield  {title} {\bibinfo {title} {Simulating
  microscopic hydrodynamic phenomena with dissipative particle dynamics},\
  }\href {https://doi.org/10.1209/0295-5075/19/3/001} {\bibfield  {journal}
  {\bibinfo  {journal} {Europhysics Letters ({EPL})}\ }\textbf {\bibinfo
  {volume} {19}},\ \bibinfo {pages} {155} (\bibinfo {year} {1992})}\BibitemShut
  {NoStop}%
\bibitem [{\citenamefont {Nikunen}\ \emph {et~al.}(2003)\citenamefont
  {Nikunen}, \citenamefont {Karttunen},\ and\ \citenamefont
  {Vattulainen}}]{nikunen2003would}%
  \BibitemOpen
  \bibfield  {author} {\bibinfo {author} {\bibfnamefont {P.}~\bibnamefont
  {Nikunen}}, \bibinfo {author} {\bibfnamefont {M.}~\bibnamefont {Karttunen}},\
  and\ \bibinfo {author} {\bibfnamefont {I.}~\bibnamefont {Vattulainen}},\
  }\bibfield  {title} {\bibinfo {title} {How would you integrate the equations
  of motion in dissipative particle dynamics simulations?},\ }\href@noop {}
  {\bibfield  {journal} {\bibinfo  {journal} {Computer physics communications}\
  }\textbf {\bibinfo {volume} {153}},\ \bibinfo {pages} {407} (\bibinfo {year}
  {2003})}\BibitemShut {NoStop}%
\bibitem [{\citenamefont {Revenga}\ \emph {et~al.}(1999)\citenamefont
  {Revenga}, \citenamefont {Zúñiga},\ and\ \citenamefont
  {Español}}]{REVENGA1999309}%
  \BibitemOpen
  \bibfield  {author} {\bibinfo {author} {\bibfnamefont {M.}~\bibnamefont
  {Revenga}}, \bibinfo {author} {\bibfnamefont {I.}~\bibnamefont {Zúñiga}},\
  and\ \bibinfo {author} {\bibfnamefont {P.}~\bibnamefont {Español}},\
  }\bibfield  {title} {\bibinfo {title} {Boundary conditions in dissipative
  particle dynamics},\ }\href
  {https://doi.org/https://doi.org/10.1016/S0010-4655(99)00341-0} {\bibfield
  {journal} {\bibinfo  {journal} {Computer Physics Communications}\ }\textbf
  {\bibinfo {volume} {121-122}},\ \bibinfo {pages} {309} (\bibinfo {year}
  {1999})},\ \bibinfo {note} {proceedings of the Europhysics Conference on
  Computational Physics CCP 1998}\BibitemShut {NoStop}%
\bibitem [{\citenamefont {McWhirter}\ \emph {et~al.}(2009)\citenamefont
  {McWhirter}, \citenamefont {Noguchi},\ and\ \citenamefont
  {Gompper}}]{mcwhirter2009flow}%
  \BibitemOpen
  \bibfield  {author} {\bibinfo {author} {\bibfnamefont {J.~L.}\ \bibnamefont
  {McWhirter}}, \bibinfo {author} {\bibfnamefont {H.}~\bibnamefont {Noguchi}},\
  and\ \bibinfo {author} {\bibfnamefont {G.}~\bibnamefont {Gompper}},\
  }\bibfield  {title} {\bibinfo {title} {Flow-induced clustering and alignment
  of vesicles and red blood cells in microcapillaries},\ }\href@noop {}
  {\bibfield  {journal} {\bibinfo  {journal} {Proceedings of the National
  Academy of Sciences}\ }\textbf {\bibinfo {volume} {106}},\ \bibinfo {pages}
  {6039} (\bibinfo {year} {2009})}\BibitemShut {NoStop}%
\bibitem [{\citenamefont {Fedosov}\ \emph {et~al.}(2011)\citenamefont
  {Fedosov}, \citenamefont {Lei}, \citenamefont {Caswell}, \citenamefont
  {Suresh},\ and\ \citenamefont {Karniadakis}}]{fedosov2011multiscale}%
  \BibitemOpen
  \bibfield  {author} {\bibinfo {author} {\bibfnamefont {D.~A.}\ \bibnamefont
  {Fedosov}}, \bibinfo {author} {\bibfnamefont {H.}~\bibnamefont {Lei}},
  \bibinfo {author} {\bibfnamefont {B.}~\bibnamefont {Caswell}}, \bibinfo
  {author} {\bibfnamefont {S.}~\bibnamefont {Suresh}},\ and\ \bibinfo {author}
  {\bibfnamefont {G.~E.}\ \bibnamefont {Karniadakis}},\ }\bibfield  {title}
  {\bibinfo {title} {Multiscale modeling of red blood cell mechanics and blood
  flow in malaria},\ }\href@noop {} {\bibfield  {journal} {\bibinfo  {journal}
  {PLoS Comput. Biol}\ }\textbf {\bibinfo {volume} {7}},\ \bibinfo {pages}
  {e1002270} (\bibinfo {year} {2011})}\BibitemShut {NoStop}%
\bibitem [{\citenamefont {Kenner}(1989)}]{kenner1989measurement}%
  \BibitemOpen
  \bibfield  {author} {\bibinfo {author} {\bibfnamefont {T.}~\bibnamefont
  {Kenner}},\ }\bibfield  {title} {\bibinfo {title} {The measurement of blood
  density and its meaning},\ }\href@noop {} {\bibfield  {journal} {\bibinfo
  {journal} {Basic research in cardiology}\ }\textbf {\bibinfo {volume} {84}},\
  \bibinfo {pages} {111} (\bibinfo {year} {1989})}\BibitemShut {NoStop}%
\bibitem [{\citenamefont {Grover}\ \emph {et~al.}(2011)\citenamefont {Grover},
  \citenamefont {Bryan}, \citenamefont {Diez-Silva}, \citenamefont {Suresh},
  \citenamefont {Higgins},\ and\ \citenamefont
  {Manalis}}]{grover2011measuring}%
  \BibitemOpen
  \bibfield  {author} {\bibinfo {author} {\bibfnamefont {W.~H.}\ \bibnamefont
  {Grover}}, \bibinfo {author} {\bibfnamefont {A.~K.}\ \bibnamefont {Bryan}},
  \bibinfo {author} {\bibfnamefont {M.}~\bibnamefont {Diez-Silva}}, \bibinfo
  {author} {\bibfnamefont {S.}~\bibnamefont {Suresh}}, \bibinfo {author}
  {\bibfnamefont {J.~M.}\ \bibnamefont {Higgins}},\ and\ \bibinfo {author}
  {\bibfnamefont {S.~R.}\ \bibnamefont {Manalis}},\ }\bibfield  {title}
  {\bibinfo {title} {Measuring single-cell density},\ }\href@noop {} {\bibfield
   {journal} {\bibinfo  {journal} {Proceedings of the National Academy of
  Sciences}\ }\textbf {\bibinfo {volume} {108}},\ \bibinfo {pages} {10992}
  (\bibinfo {year} {2011})}\BibitemShut {NoStop}%
\bibitem [{\citenamefont {Bryan}\ \emph {et~al.}(2014)\citenamefont {Bryan},
  \citenamefont {Hecht}, \citenamefont {Shen}, \citenamefont {Payer},
  \citenamefont {Grover},\ and\ \citenamefont {Manalis}}]{bryan2014measuring}%
  \BibitemOpen
  \bibfield  {author} {\bibinfo {author} {\bibfnamefont {A.~K.}\ \bibnamefont
  {Bryan}}, \bibinfo {author} {\bibfnamefont {V.~C.}\ \bibnamefont {Hecht}},
  \bibinfo {author} {\bibfnamefont {W.}~\bibnamefont {Shen}}, \bibinfo {author}
  {\bibfnamefont {K.}~\bibnamefont {Payer}}, \bibinfo {author} {\bibfnamefont
  {W.~H.}\ \bibnamefont {Grover}},\ and\ \bibinfo {author} {\bibfnamefont
  {S.~R.}\ \bibnamefont {Manalis}},\ }\bibfield  {title} {\bibinfo {title}
  {Measuring single cell mass, volume, and density with dual suspended
  microchannel resonators},\ }\href@noop {} {\bibfield  {journal} {\bibinfo
  {journal} {Lab on a Chip}\ }\textbf {\bibinfo {volume} {14}},\ \bibinfo
  {pages} {569} (\bibinfo {year} {2014})}\BibitemShut {NoStop}%
\bibitem [{\citenamefont {Backer}\ \emph {et~al.}(2005)\citenamefont {Backer},
  \citenamefont {Lowe}, \citenamefont {Hoefsloot},\ and\ \citenamefont
  {Iedema}}]{backer2005poiseuille}%
  \BibitemOpen
  \bibfield  {author} {\bibinfo {author} {\bibfnamefont {J.}~\bibnamefont
  {Backer}}, \bibinfo {author} {\bibfnamefont {C.}~\bibnamefont {Lowe}},
  \bibinfo {author} {\bibfnamefont {H.}~\bibnamefont {Hoefsloot}},\ and\
  \bibinfo {author} {\bibfnamefont {P.}~\bibnamefont {Iedema}},\ }\bibfield
  {title} {\bibinfo {title} {Poiseuille flow to measure the viscosity of
  particle model fluids},\ }\href@noop {} {\bibfield  {journal} {\bibinfo
  {journal} {The Journal of chemical physics}\ }\textbf {\bibinfo {volume}
  {122}},\ \bibinfo {pages} {154503} (\bibinfo {year} {2005})}\BibitemShut
  {NoStop}%
\bibitem [{\citenamefont {G.~Villlalobos}(2025)}]{tcruzidataset}%
  \BibitemOpen
  \bibfield  {author} {\bibinfo {author} {\bibfnamefont {A.~C.}\ \bibnamefont
  {G.~Villlalobos}},\ }\href {https://doi.org/10.6084/m9.figshare.30403102}
  {\emph {\bibinfo {title} {Dataset for the paper "Mechanical Evidence of the
  imposibility of directed motion ofTrypanosoma cruzi towards preferred organs
  in the HumanBody, a simulation 2D model within a laminar flow"}}},\ \bibinfo
  {type} {Tech. Rep.}\ (\bibinfo  {institution} {ESAP, UJTL},\ \bibinfo {year}
  {2025})\BibitemShut {NoStop}%
\bibitem [{\citenamefont {Lee}\ and\ \citenamefont
  {Chen}(2002)}]{lee2002numerical}%
  \BibitemOpen
  \bibfield  {author} {\bibinfo {author} {\bibfnamefont {D.}~\bibnamefont
  {Lee}}\ and\ \bibinfo {author} {\bibfnamefont {J.}~\bibnamefont {Chen}},\
  }\bibfield  {title} {\bibinfo {title} {Numerical simulation of steady flow
  fields in a model of abdominal aorta with its peripheral branches},\
  }\href@noop {} {\bibfield  {journal} {\bibinfo  {journal} {Journal of
  Biomechanics}\ }\textbf {\bibinfo {volume} {35}},\ \bibinfo {pages} {1115}
  (\bibinfo {year} {2002})}\BibitemShut {NoStop}%
\bibitem [{\citenamefont {Wootton}\ and\ \citenamefont
  {Ku}(1999)}]{wootton1999fluid}%
  \BibitemOpen
  \bibfield  {author} {\bibinfo {author} {\bibfnamefont {D.~M.}\ \bibnamefont
  {Wootton}}\ and\ \bibinfo {author} {\bibfnamefont {D.~N.}\ \bibnamefont
  {Ku}},\ }\bibfield  {title} {\bibinfo {title} {Fluid mechanics of vascular
  systems, diseases, and thrombosis},\ }\href@noop {} {\bibfield  {journal}
  {\bibinfo  {journal} {Annual review of biomedical engineering}\ }\textbf
  {\bibinfo {volume} {1}},\ \bibinfo {pages} {299} (\bibinfo {year}
  {1999})}\BibitemShut {NoStop}%
\bibitem [{\citenamefont {Ku}(1997)}]{ku1997blood}%
  \BibitemOpen
  \bibfield  {author} {\bibinfo {author} {\bibfnamefont {D.~N.}\ \bibnamefont
  {Ku}},\ }\bibfield  {title} {\bibinfo {title} {Blood flow in arteries},\
  }\href@noop {} {\bibfield  {journal} {\bibinfo  {journal} {Annual Review of
  Fluid Mechanics}\ }\textbf {\bibinfo {volume} {29}},\ \bibinfo {pages} {399}
  (\bibinfo {year} {1997})}\BibitemShut {NoStop}%
\bibitem [{\citenamefont {Uppaluri}(2011)}]{uppaluri2011unicellular}%
  \BibitemOpen
  \bibfield  {author} {\bibinfo {author} {\bibfnamefont {S.}~\bibnamefont
  {Uppaluri}},\ }\bibfield  {title} {\bibinfo {title} {Unicellular parasite
  motility: a quantitative perspective},\ }\href@noop {} {\  (\bibinfo {year}
  {2011})}\BibitemShut {NoStop}%
\bibitem [{\citenamefont {Molnar}\ and\ \citenamefont
  {Labouesse}(2021)}]{molnar2021plastic}%
  \BibitemOpen
  \bibfield  {author} {\bibinfo {author} {\bibfnamefont {K.}~\bibnamefont
  {Molnar}}\ and\ \bibinfo {author} {\bibfnamefont {M.}~\bibnamefont
  {Labouesse}},\ }\bibfield  {title} {\bibinfo {title} {The plastic cell:
  mechanical deformation of cells and tissues},\ }\href@noop {} {\bibfield
  {journal} {\bibinfo  {journal} {Open Biology}\ }\textbf {\bibinfo {volume}
  {11}},\ \bibinfo {pages} {210006} (\bibinfo {year} {2021})}\BibitemShut
  {NoStop}%
\end{thebibliography}%

\end{document}